%% file: www2004.tex
\documentclass[11pt,letterpaper]{article}
\usepackage{fullpage,times}
\input{psfig-dvips}

\newif\ifpdf
\ifx\pdfoutput\undefined
  \pdffalse
\else
  \pdfoutput=1
  \pdftrue
\fi

\ifpdf
  \usepackage[pdftex]{graphicx}
  \usepackage[pdftex]{color}
  \DeclareGraphicsExtensions{.pdf,.png,.jpg}
\else
  \usepackage[dvips]{graphicx}
  \usepackage[dvips]{color}
  \DeclareGraphicsExtensions{.eps,.epsi,.epsf,.ps}
\fi

\begin{document}

\title{Staging Transformations for\\
Multimodal Web Interaction Management}

\author{Michael Narayan, Chris Williams, Saverio Perugini, and Naren Ramakrishnan\\
Department of Computer Science\\
Virginia Tech, Blacksburg, VA 24061, USA\\
E-mail: \{mnarayan, chwilli4, saverio, naren\}@vt.edu\\
Project website: http://pipe.cs.vt.edu}
\maketitle

\begin{abstract}
Multimodal interfaces are becoming increasingly ubiquitous with the advent of 
mobile devices, accessibility considerations, and novel software technologies 
that combine diverse interaction media. In addition to improving access and delivery
capabilities, such interfaces enable flexible and personalized dialogs 
with websites, much like a conversation between humans. In this paper, we 
present a software framework for multimodal web interaction management that supports
mixed-initiative dialogs between users and websites.  A mixed-initiative dialog is 
one where the user and the website take turns changing the flow of interaction. 
The framework supports the functional specification and realization of such
dialogs using staging transformations -- a theory for representing and reasoning 
about dialogs based on partial input. It supports multiple interaction interfaces,
and offers sessioning, caching, and co-ordination functions
through the use of an interaction manager. Two case studies are presented 
to illustrate the promise of this approach.
\end{abstract}

\thispagestyle{empty}

\vspace{-0.1in}
\paragraph{Categories and Subject Descriptors:}
H.5.4 [{\bf Hypertext/Hypermedia}]: Navigation;
H.5.2 [{\bf User Interfaces}]: Interaction Styles;
F.3.2 [{\bf Semantics of Programming Languages}]: Partial Evaluation

\paragraph{Terms:}
interaction management in web applications, multimodal interfaces, novel
browsing paradigms.

\vspace{-0.1in}
\paragraph{Keywords:}
program transformations, partial evaluation, mixed-initiative interaction,
out-of-turn interaction, web dialogs.

\newpage

\section{Introduction}
Web interaction management is a well-studied topic and has produced many innovative
solutions to support contextual interactions with websites~\cite{mawl,bigwig}. Today's 
web systems feature
a diverse range of interactive functionality, from preserving state across sessions
(e.g., shopping carts at amazon.com) to gracefully accommodating users' interruptive
activities such as pressing back buttons and cloning windows 
(\cite{continuationsWebServers}; e.g., in form-based
services). With the shift of web access from the desktop to mobile devices such as PDAs,
tablet PCs, and 3G phones~\cite{know-encap-pervas,whole-parts,DetectingWebPage,pers-pocket-dirs,myers-ieeecomp},
and the advent of novel multimodal interfaces,
the importance of interaction management has only become accentuated.
We posit that the logical culmination of interaction management research
will be to enable flexible and personalized dialogs with websites~\cite{piis}, much like a 
conversation between humans.

Viewing web interactions as dialogs can be very instructive, and suggests
useful metaphors. Imagine the interaction between a user and a website to
be a conversation between two participants.
The conversation typically begins as {\it user-initiated} (since the user chose 
to visit the site). The site then presents a choice of hyperlinks from 
which the user is expected to make a selection. This step of the 
conversation is {\it site-initiated} since by clicking on a(ny) hyperlink, the 
user will be responding to the 
initiative taken by the website. After following a link, the user might 
decide to pursue further links (the initiative thus residing
with the site), or press the back button (hence retaking the
initiative). Such an interaction where the user and the site exchange initiative
is called a {\it mixed-initiative} dialog.

Mixed-initiative interaction (MII) has been well studied by the 
speech interfaces, artificial intelligence planning, and discourse analysis 
communities~\cite{MII-UR} but has only recently begun to be investigated in
web interactions~\cite{MII-ITPro}. This is because the mechanisms for the user
to take the initiative in web interactions used to be limited. With the emergence
of the multimodal web and the availability of multiple paradigms for interaction,
these opportunities are now expanding; thus setting the stage for MII to
play a prominent role in web interactions.

\begin{figure}
\centering
\vspace{-0.2in}
\begin{tabular}{c}
\includegraphics[width=8.4cm]{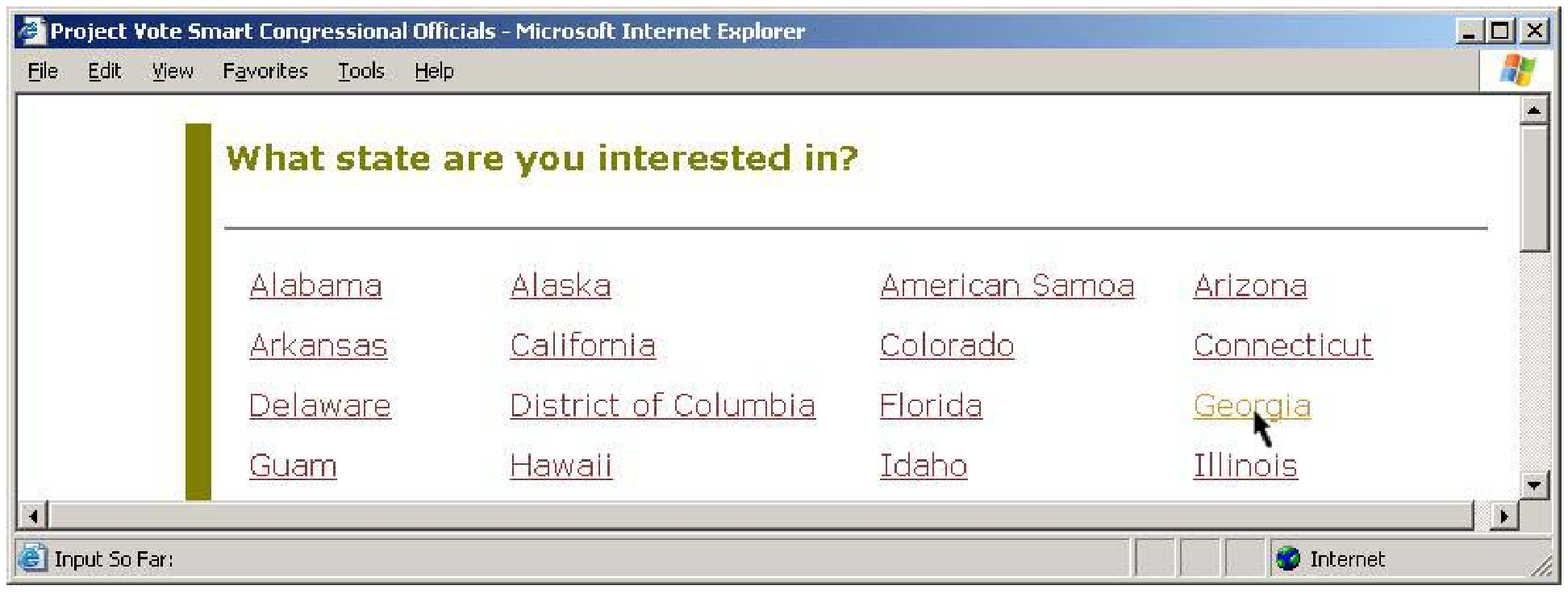}\\
$\mathbf{\Downarrow}$\\
\includegraphics[width=8.4cm]{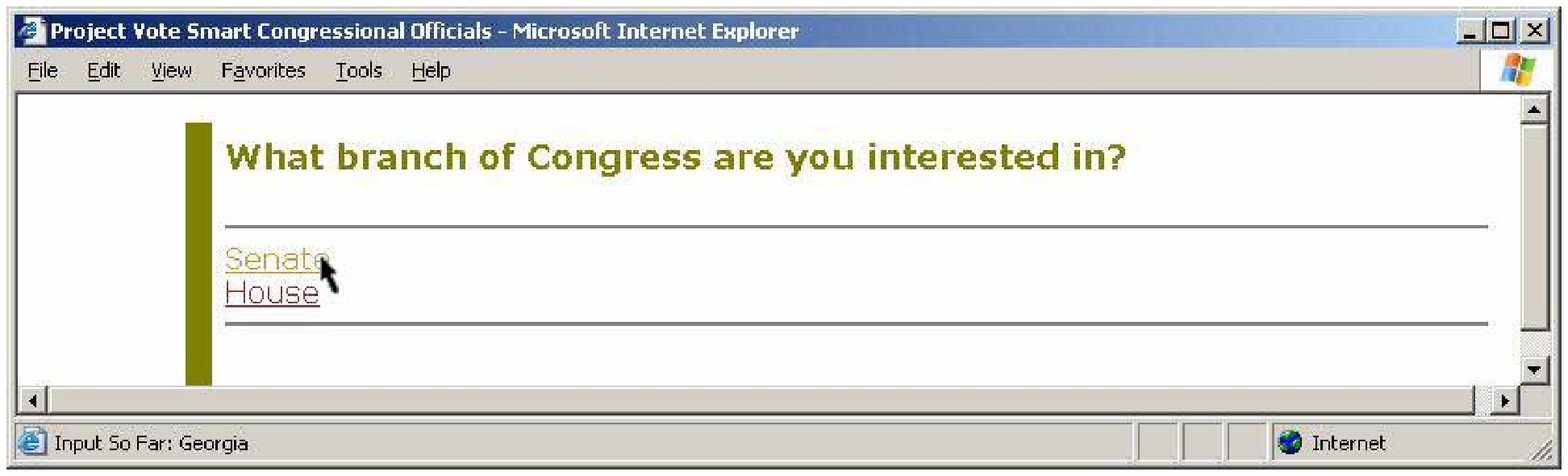}\\
$\mathbf{\Downarrow}$\\
\includegraphics[width=8.4cm]{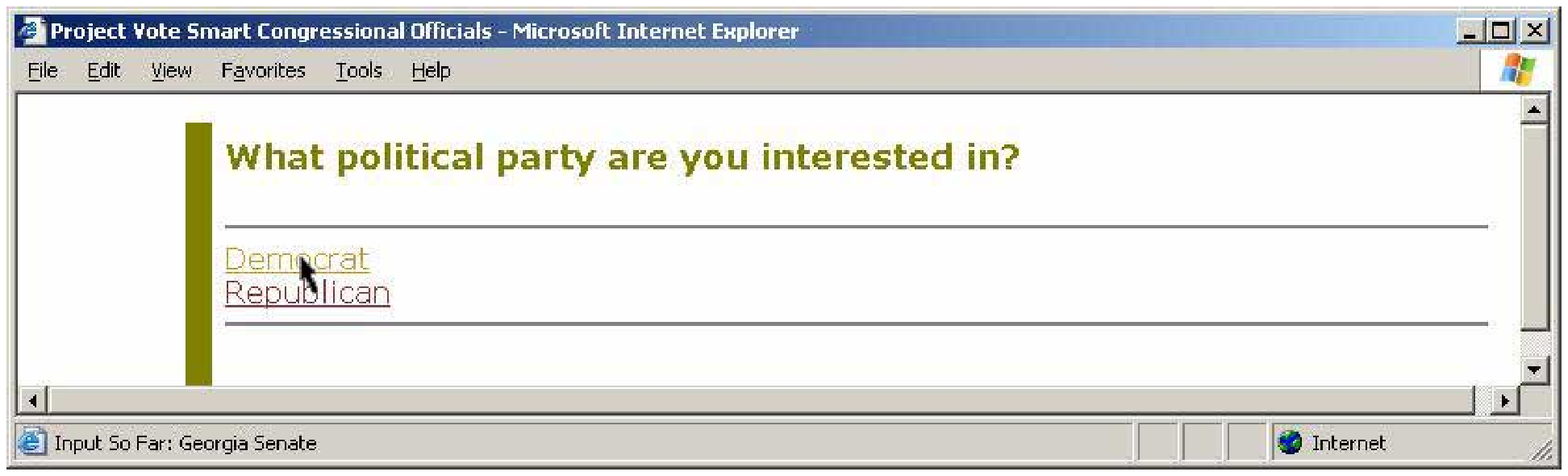}\\
$\mathbf{\Downarrow}$\\
\includegraphics[width=8.4cm]{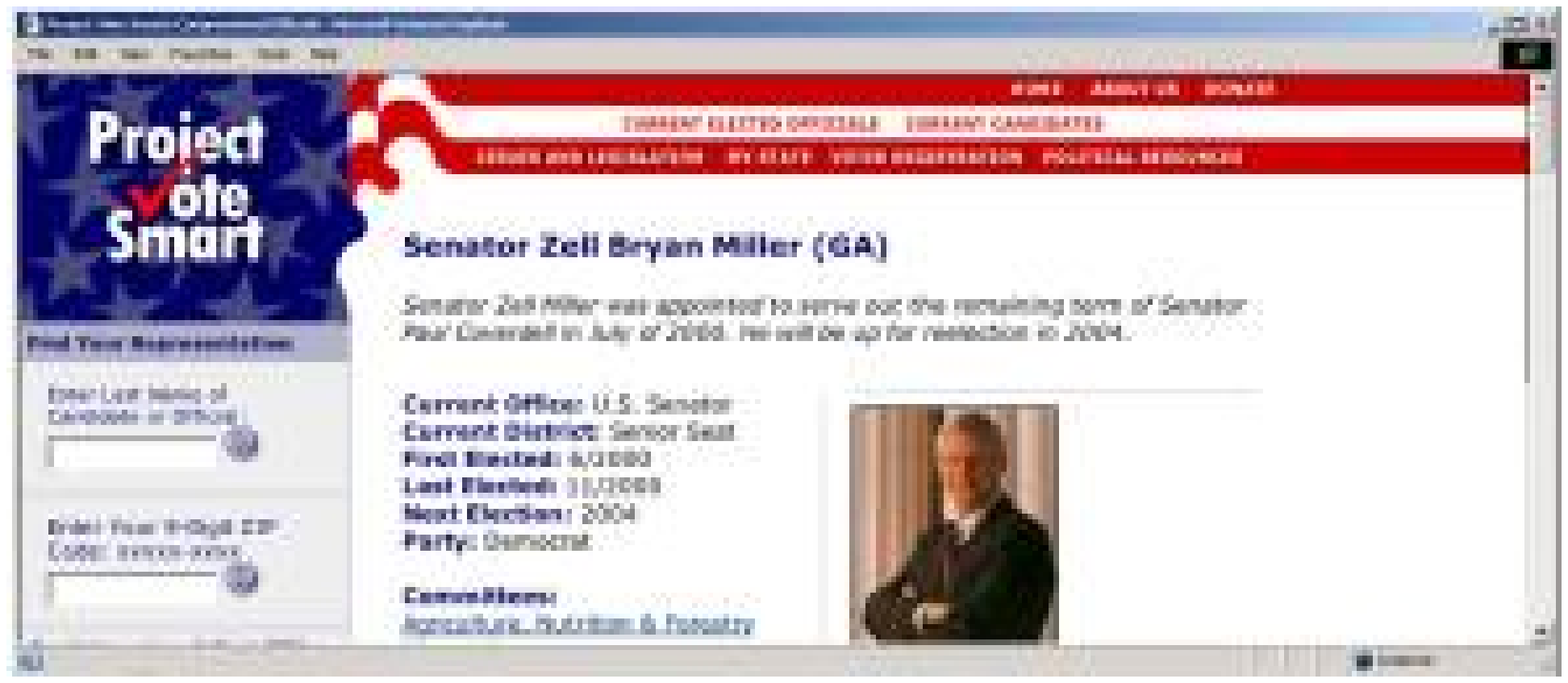}\\
\end{tabular}
\centering
\caption{A site-initiated dialog in a US congressional site to reach the webpage
of the Democratic Senator from Georgia.}
\label{votesmart-it}
\end{figure}

The primary development we are alluding to is, of course, the emergence of
the speech-enabled web~\cite{introCACMSep2000}; technologies such as SALT (Speech Application
Language Tags;~\cite{SALT}) and X+V (XHTML plus Voice;~\cite{xandv}) are ushering in documents that
can talk and listen rather than passively display content.
The maturing of commercial speech recognition
engines~\cite{speechRec}, the naturalness of speech for
conversational interaction, and its role 
in improving accessibility (i.e.,
for visually impaired people) have
been key factors in the emergence of this niche segment of multimodal browsing.

Speech is commonly perceived as merely as a vocal substitute to
hyperlink access (i.e., `say' a hyperlink label instead
of clicking on it)~\cite{WebViews,AVoN}. Our approach, on the other hand, is to think
of speech as a way for the user to take the initiative in web interactions, and hence
{\it augment} hyperlink usage. While
hyperlink access must, by definition, be responsive to the initiative taken by the
site, speech interaction can be used to either respond to or take the initiative. 
Our view of the multimodal web is hence one that enables flexible dialogs, with rich
opportunities for mixed-initiative interaction. We focus our studies here on speech
and hyperlink as the modes of interaction although our framework will apply more generally
to new input mechanisms.

\subsection*{Motivating Example}
\begin{figure}
\centering
\vspace{-0.2in}
\begin{tabular}{c}
\includegraphics[width=8.4cm]{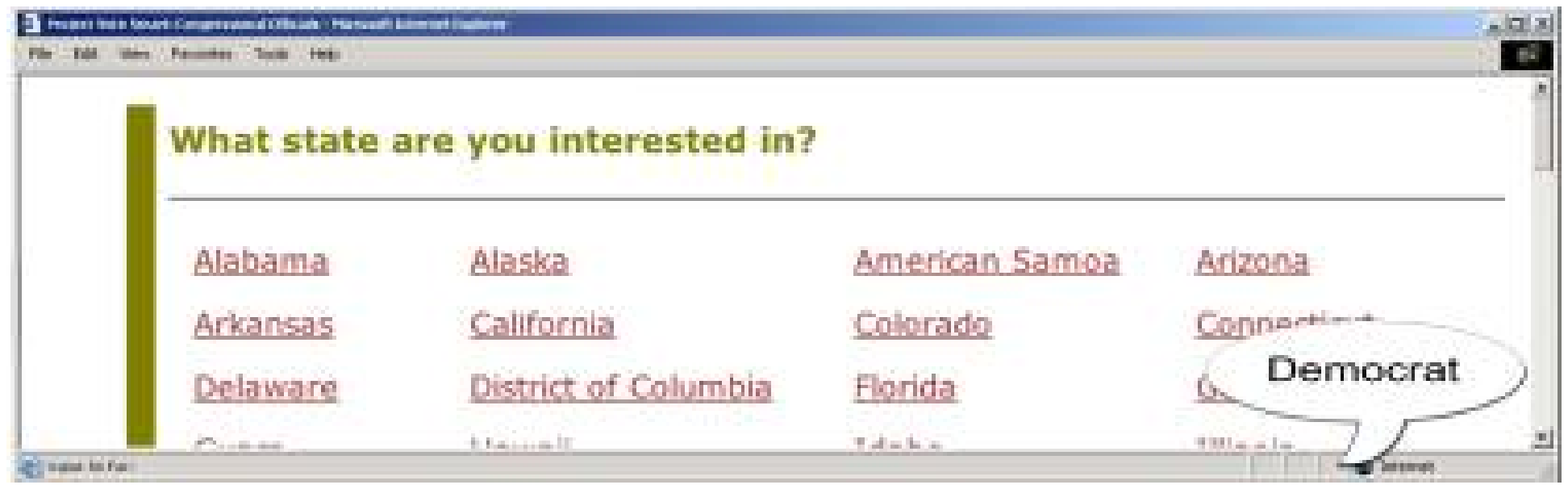}\\
$\mathbf{\Downarrow}$\\
\includegraphics[width=8.4cm]{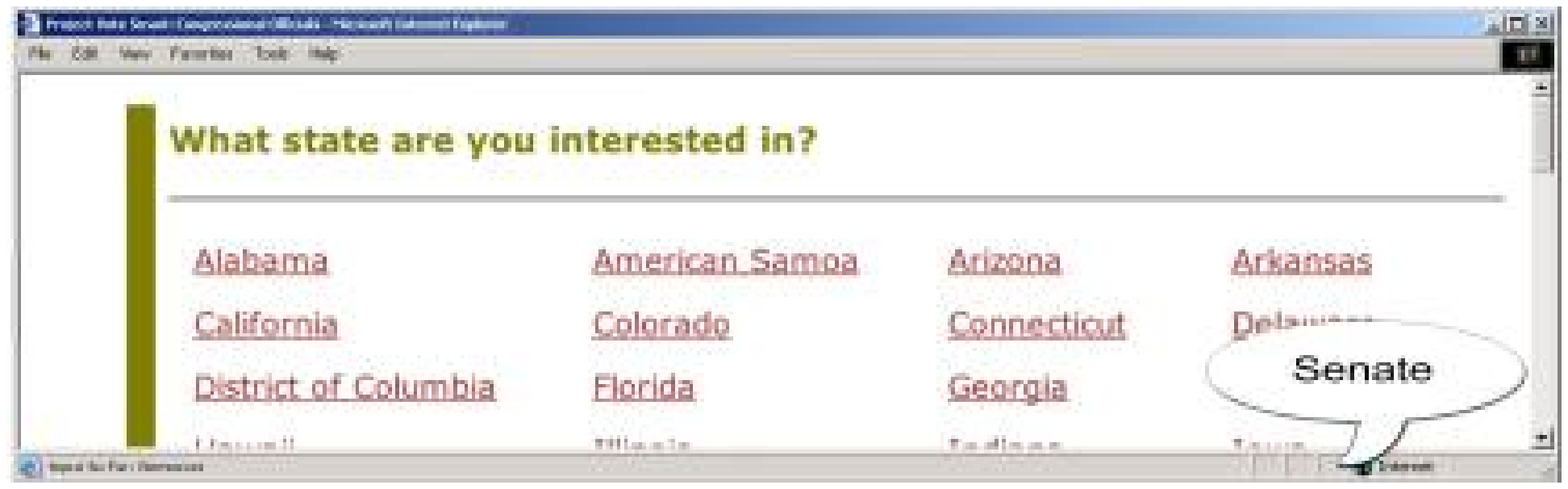}\\
$\mathbf{\Downarrow}$\\
\includegraphics[width=8.4cm]{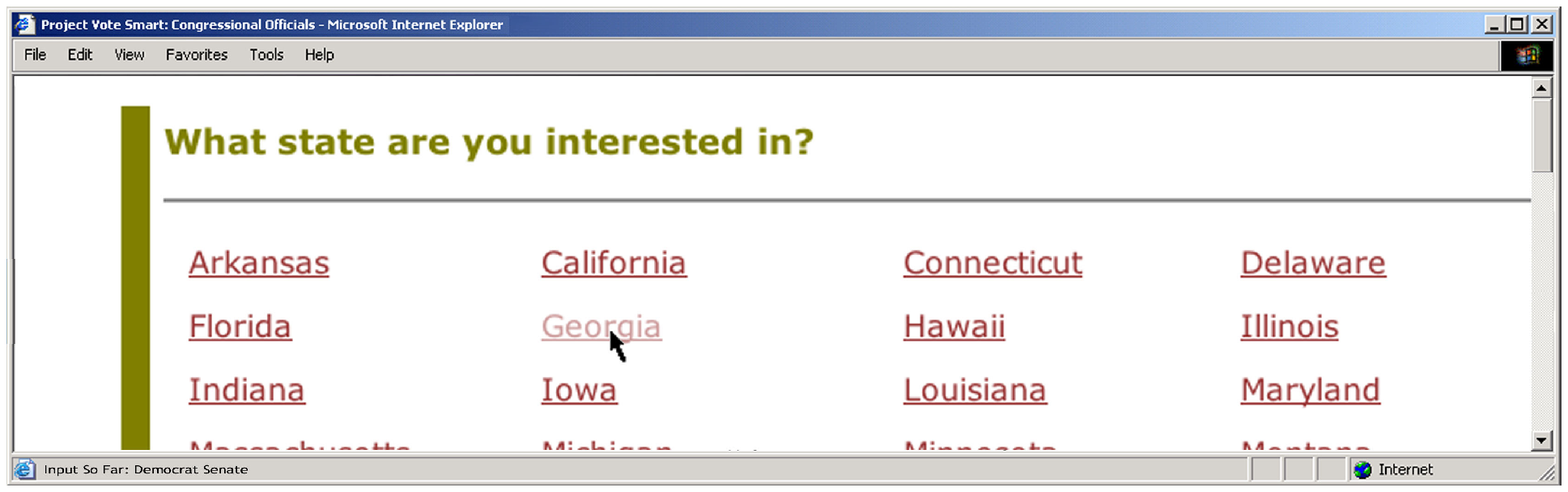}\\
$\mathbf{\Downarrow}$\\
\includegraphics[width=8.4cm]{figs/votesmartFinal.eps}\\
\end{tabular}
\centering
\caption{A mixed-initiative dialog  in a US congressional site to reach the webpage
of the Democratic Senator from Georgia.}
\label{votesmart-oot}
\vspace{-0.1in}
\end{figure}

Project Vote Smart (vote-smart.org) is a hierarchically organized site for 
information about US Congressional officials. The first level in this 
site corresponds
to choice of state, the second level corresponds to branch of congress, 
the third level for party, and so on\footnote{Since this writing,
the Vote Smart site has been 
restructured into a flat faceted classification; nevertheless the observations and
central ideas of this paper still apply.}.
Fig.~\ref{votesmart-it} and Fig.~\ref{votesmart-oot} depict two
dialogs of interaction with Vote Smart; while both culminate in the webpage 
of Senator Miller they involve fundamentally different interaction sequences. 

The dialog in Fig.~\ref{votesmart-it} is site-initiated since
the user progressively
clicks on presented hyperlinks (Georgia: Senate: Democrat) to specify values 
for relevant
politician attributes. All such browsing interactions are responsive
to the current solicitation; we sometimes refer to such an interaction sequence
as an {\it in-turn} sequence. 

Fig.~\ref{votesmart-oot} describes a web session with
capabilities for speech input. At any stage in the dialog the user has the option
of either pursuing a hyperlink or speaking some utterance. In the top
panel of Fig.~\ref{votesmart-oot}, when the website solicits for choice of
state, the user responds with his choice of party (Democrat) instead, by speaking
out-of-turn. The user has hence taken the initiative. Such an interaction is 
called an {\it out-of-turn} interaction since the user is specifying 
party information at the first level (when it is normally specified at 
the third level). As a result of this out-of-turn input, 
states which do not have any Democratic politicians are pruned out (e.g., Alaska).
Notice that the website continues to solicit for state, since it remains
unspecified. In conversational terms, we say that the website has reclaimed 
the initiative, and is repeating its prompt for input.
At this point the user again speaks out-of-turn, this time with branch 
information (Senate). More states are pruned out (e.g., Alabama and
Arizona). 
After this step, the user reverts back to in-turn
mode, and responds to the site's solicitation by
clicking on the `Georgia' hyperlink; leading again, directly, to Senator Miller's webpage.
While there can be only one purely in-turn interaction sequence to arrive at 
a webpage, there could be several involving out-of-turn interactions. Notice
that in a single out-of-turn interaction, more than one aspect could be specified.

The above example is admittedly a simple form of mixing initiative (for more
complicated flavors, see~\cite{MII-UR,SMARTEdit}) but provides a 
powerful mechanism to interact with websites. In particular, the capability
for out-of-turn interaction obviates the need to express interaction sequences
directly in the browsing structure (i.e., as in parallel faceted browsing
classifications~\cite{Flamenco}). From an information seeking standpoint, out-of-turn 
interaction is a flexible and unintrusive way to bridge any mental mismatch 
between users and websites,
by increasing the opportunities for communicating partial input. 

\subsubsection*{Contributions of this Paper}
The goal of this paper is to flexibly support mixed-initiative 
dialogs with multimodal websites; we present a cross-platform web service architecture that 
factors multimodal interaction management into the three facets of interaction 
interfaces, a transformation engine, and an interaction manager.
Our main contributions can be summarized alongside each of these facets:

\begin{enumerate}
\itemsep=0.3pt
\parsep=5pt
\item Interaction interfaces support hyperlink and speech modes of input
and allow users to employ them uniformly in a multimodal session;
\item The transformation engine uses 
{\it staging transformations}~\cite{vt-staging} --- a functional
approach to dialog management --- to specify, reason about, and implement web 
dialogs without side-effects. In particular, staging transformations allow us 
to automatically enable existing sites for multimodal interaction, without
manual re-engineering;
\item The transformation engine further uses the staging notation to implicitly
capture state
information in a web dialog, seamlessly supporting save, restore, and caching functionalities;
\item Both categorical and non-categorical modeling is supported in the
staging transformations framework, allowing it
to be applicable to the vast majority of existing sites.
\item The interaction manager built around these concepts helps realize
sessioning, caching, and co-ordination functionality; 
And, most importantly,
\item the integration of the above ideas  
supports both in-turn and out-of-turn modes of
interaction in a unified manner, obviating the need to ever distinguish between
them.
\end{enumerate}
Finally, we argue that the factoring presented here can be easily extended to support novel
modalities of interaction in emerging application domains.

\section{Basic Approach}
To understand the staging transformations philosophy, let us revisit the dialogs
of Figs.~\ref{votesmart-it} and~\ref{votesmart-oot}. We begin with the provocative
observation that both in-turn and out-of-turn interactions
can be supported by the same dialog programming model!

\begin{figure}
\centering
\vspace{-0.2in}
\begin{tabular}{c}
\includegraphics[width=17cm]{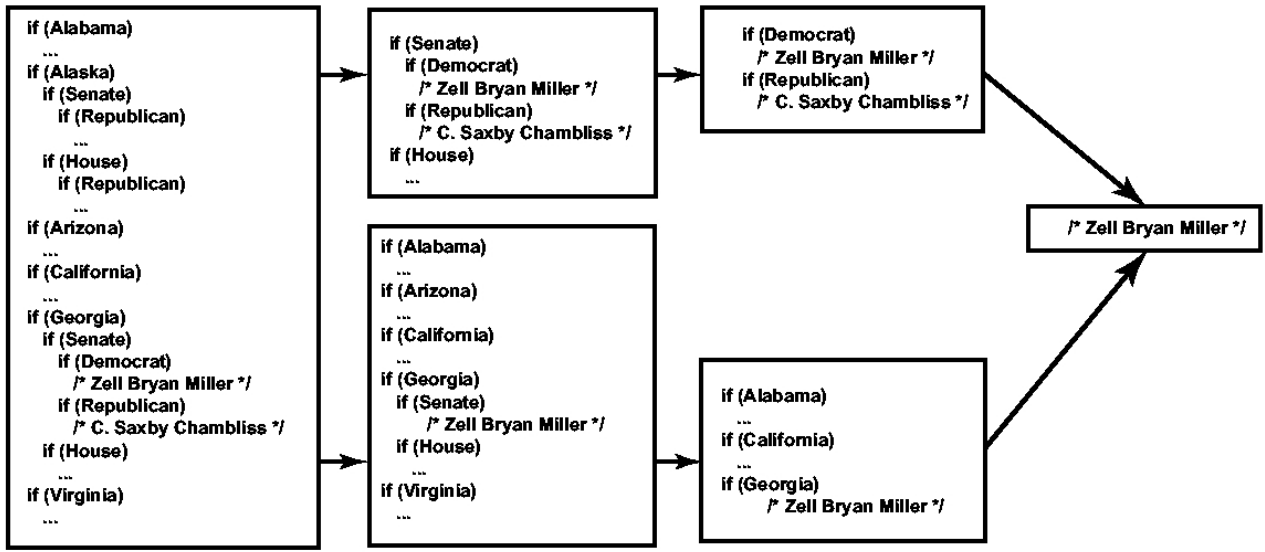}
\end{tabular}
\centering
\caption{Staging dialogs using program transformations. The top series
of transformations mimic an in-turn dialog with the user specifying
(Georgia: Senate: Democrat), in that order. The bottom series of transformations
correspond to a mixed-initiative dialog where the user 
specifies (Democrat: Senator: Georgia), in that order.}
\label{sens-staged}
\end{figure}

To see how, it is helpful to think of modeling the Vote Smart website as the program 
of Fig.~\ref{sens-staged} (left) where the nesting of conditionals reflects the hierarchical
hyperlink structure and each program variable denotes a hyperlink label. For an in-turn 
sequence, the top series of transformations in Fig.~\ref{sens-staged}
depicts what we want to happen. 
For the interaction of Fig.~\ref{votesmart-oot}, the bottom series of
transformations depicts what we want to happen. Notice that both sequences start and end with the same representation, 
but take different paths.

The first sequence of transformations corresponds to {\it interpreting} the program in the order in which
it is written, i.e., when the user clicks on `Georgia,' that variable is set to one and all other
state variables (e.g., `Alabama') are set to zero, and the program is interpreted. This leads
to a simplified program that now solicits for branch of congress. The second sequence of
transformations involves `jumping ahead' to nested program segments and simplifying them even before
outer portions are evaluated. Such a non-sequential evaluation
is well known in the programming languages
literature to be {\it partial evaluation} (\cite{introPartialEvaluation};
see Fig.~\ref{pe-example}), a technique for 
specializing programs given some (but not all)
of their input. Thus, when the user says `Democrat' out-of-turn, the program is partially evaluated
with this variable set to one (and `Republican' set to zero). The simplified program continues to solicit
for state at the top level, but some states are now removed since the
corresponding program segments involve dead-ends.
Notice that since partial evaluation can be used for interpretation, it 
can support the first interaction sequence as well. 

\begin{figure}
\centering
\begin{tabular}{|l|l|}\hline
{\tt int pow(int base, int exponent) \{} & {\tt int pow2(int base) \{} \\
\,\,\,\,\,{\tt int prod = 1;} & \,\,\,\,\,{\tt return (base * base);} \\
\,\,\,\,\,{\tt for (int i = 0; i < exponent; i++)} &  \} \\
\,\,\,\,\,\,\,\,\,\,{\tt prod = prod * base;} & \\
\,\,\,\,\,{\tt return prod;} & \\
\} & \\
\hline
\end{tabular}
\caption{Illustration of the partial evaluation technique.
A general purpose {\tt pow}er function written in C (left) and
its specialized version (with {\tt exponent} statically set to 2) to handle
squares (right). Automatic partial evaluators (e.g., C-Mix) use techniques 
such as loop unrolling and copy propagation to specialize given programs.}
\label{pe-example}
\end{figure}

This is the essence of our staging transformations framework: writing a program to model
the structure of the dialog and using a program transformer to {\it stage} it. In addition,
we must adopt a way to map users' partial inputs to assignments of values to
program variables.

\begin{figure}
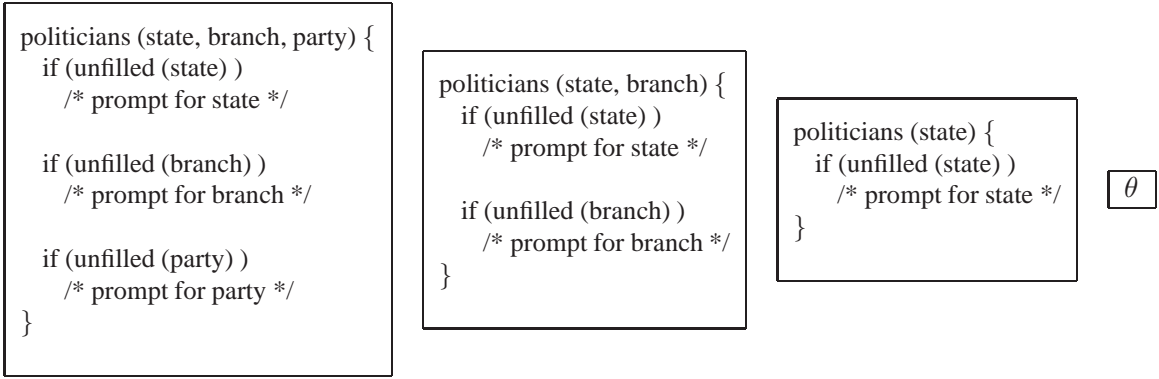

\centering
\begin{tabular}{cccc}
\begin{tabular}{|p{1.4in}|} \hline
\vspace{-0.2in}
\small
\begin{tabbing}
politicians (state, branch, party) \{\\
\,\,\,\,\,if (unfilled (state) ) \\
\,\,\,\,\,\,\,\,\,\,/* prompt for state */\\
\,\,\,\,\,\, \\
\,\,\,\,\,if (unfilled (branch) ) \\
\,\,\,\,\,\,\,\,\,\,/* prompt for branch */\\
\,\,\,\,\,\, \\
\,\,\,\,\,if (unfilled (party) ) \\
\,\,\,\,\,\,\,\,\,\,/* prompt for party */\\
\}
\end{tabbing}
\vspace{-0.15in}
\\\hline
\end{tabular}
&
\begin{tabular}{|p{1.4in}|} \hline
\vspace{-0.2in}
\small
\begin{tabbing}
politicians (state, branch) \{\\
\,\,\,\,\,if (unfilled (state) ) \\
\,\,\,\,\,\,\,\,\,\,/* prompt for state */\\
\,\,\,\,\,\, \\
\,\,\,\,\,if (unfilled (branch) ) \\
\,\,\,\,\,\,\,\,\,\,/* prompt for branch */\\
\}
\end{tabbing}
\vspace{-0.15in}
\\\hline
\end{tabular}
&
\begin{tabular}{|p{1.4in}|} \hline
\vspace{-0.2in}
\small
\begin{tabbing}
politicians (state) \{\\
\,\,\,\,\,if (unfilled (state) ) \\
\,\,\,\,\,\,\,\,\,\,/* prompt for state */\\
\}
\end{tabbing}
\vspace{-0.15in}
\\\hline
\end{tabular}
&
\begin{tabular}{|p{0.08in}|} \hline
$\theta$
\vspace{-0.2in}
\\\hline
\end{tabular}
\end{tabular}
\caption{Representing the politicians site using categorical variables and
staging the dialog of Fig.~\ref{votesmart-oot}
using partial evaluation.
$\theta$ denotes the empty dialog.}
\label{sens-alternate}
\vspace{-0.2in}
\end{figure}

There are many ways to model dialog structure
as programs and the choice of representation is dependent 
on both the structural characteristics of the website and the interaction 
scenarios that must be supported. For instance, we can take advantage of the 
levelwise property of the politicians website and
model the dialog as shown in Fig.~\ref{sens-alternate}. In this 
representation, the user's input is captured as assignment of values to 
only three categorical variables, whereas the representation of 
Fig.~\ref{sens-staged} used many more variables, but all boolean. 

The representation of Fig.~\ref{sens-staged}
is suitable in a setting where we would like to provide dynamic feedback
to the user during the course of an interaction. For instance, even after just 
supplying `Democrat' at the outset,
the user is provided feedback such as the fact that there are 
no Democratic politicians (senators or otherwise) in Alaska. The representation 
of Fig.~\ref{sens-alternate}
does not explicitly capture such dependencies and
is more appropriate in a database-driven website where a lookup 
happens only after all relevant information
is collected. Furthermore, this representation assumes that there 
is a way to map user's inputs to
the relevant categories (e.g., when the user says `Alaska,' it must be inferred 
that he is talking about a state). There is no automatic means to obtain this 
categorical assignment by a program analysis. 
Whereas, in the representation of
Fig.~\ref{sens-staged}, we would merely need to infer that `Democrat' excludes
the possibility of being `Republican,' a feature that can be captured mechanically
through the application of the program transformation (see end of Section~\ref{michael}).

Staging transformations thus provide a functional, implementation-neutral way
to specify web dialogs; to fully realize our vision of a flexible multimodal
web interaction framework, we require:
\begin{itemize}
\itemsep=0pt
\parsep=3pt
\item A greater variety of stagers to support practical web dialogs;
\item A theory for reasoning about dialogs based on user input;
\item A robust transformation engine based on commercial web technologies;
\item Interaction interfaces for capturing and communicating user input; and
\item an interaction manager providing session management, co-ordination, and
caching functionality.
\end{itemize} 

\section{System Overview and Design}
\subsection{Staging Transformations}
\label{michael}
\subsubsection*{Dialog Notation}
We begin by introducing a notation to represent the structure 
of dialogs as well as the program transformations for staging 
them. The notation is easiest to understand when only 
categorical variables are involved. A purely in-turn sequence in the politicians
site would be denoted by:
$${\frac{I}{\textrm{state}\ \ \textrm{branch}\ \ \textrm{party}}}$$
where the $I$ indicates that an interpreter is used as the staging transformer. Similarly, 
$${\frac{PE}{\textrm{state}\ \ \textrm{branch}\ \ \textrm{party}}}$$
denotes a dialog staged with a partial evaluator (PE).
An interpreter permits only
inputs that are responsive to the current solicitation and proceeds in a strict
sequential order; it results in the most restrictive dialog.
A PE, on the other hand, allows utterances of any combination of available
input slots in the dialog. It allows all $3!$ orderings of the
politician attributes (13, if we allow multiple attributes per utterance)
to be achieved, without explicitly programming for them. 
It is the most flexible of stagers.  However, it is an all-or-nothing stager
that cannot enforce (i.e., require) a particular ordering. 
With the above notation, hence just replacing the $I$ in a dialog with a $PE$ dramatically
changes a website from one that is meant for browsing to one that supports
out-of-turn interaction!

Stagers can be composed in a hierarchical fashion to yield dialogs
comprised of smaller dialogs, or subdialogs. This allows us to make
fine-grained distinctions about the structure of dialogs and the range of
valid inputs. In this sense,
\[\frac{PE}{a\:b\:c\:d}\]
\vspace{-0.05in}
is not the same as
\vspace{-0.05in}
\[\frac{PE}{\frac{PE}{a\:b}\frac{PE}{c\:d}}\]
The former allows all $4!$ permutations of $\{a,b,c,d\}$ whereas
the latter precludes utterances such as $\prec\!c\:a\:b\:d\!\succ$.

Another useful stager is the currier (C), which corresponds to 
the standard definition of currying from
the programming languages literature. Programmatically, currying is
a specialization of a function when some partial prefix of its arguments
are known. From a dialog viewpoint, a curryer allows for
multiple utterances to be made at one time, with the restriction
that these utterances must fill in the dialog arguments in
consecutive order, starting at the beginning of the dialog.  

To support non-categorical modeling such as shown in Fig.~\ref{sens-staged}
we require the ability to model webpages with multiple links, where only
one link can be pursued at a given time. The alternator stager (A) addresses
this requirement; while it can be implemented as a program transformation, it is
not particularly interesting to look at it as such, since it has
little meaning in other contexts. 

Let us investigate how to specify a browsing interaction with the politicians website.
Since the partial inputs correspond to hyperlink labels, the dialog representation
must involve a hierarchical composition over these labels:

\[
    \frac{A}
    {\frac{I}{
	\textrm{ga}
        {\frac{A}
        {\frac{I} {
	    \textrm{s}
            {\frac{A} {
                \textrm{r}\;
                \textrm{d}}
            }}
        \frac{I} {\textrm{h} \ldots}
        }
        }}
    \frac{I}{\textrm{ak} \ldots}
    \frac{I}{\textrm{al} \ldots}
    \ldots
    }
\]

At the top-level of this dialog, there is a choice of multiple subdialogs, all of
which involve the specification of some state label. We see that from
the first page, there are links to pages for Georgia (`ga'),
Arkansas (`ak'), and so on. Notice that the order in which we list these links
does not matter since, from the viewpoint of the A stager, only one of the
corresponding subdialogs can
be entered. Looking at the link for Georgia, which is expanded in more detail,
we see that this subdialog consists of a further choice among 
senate (`s') versus house ('h) politicians, and so on. The
I's emphasize that that levels must be entered in strict sequential order, reinforcing
the browsing paradigm. If we replace the Is with PEs, we will
effectively specify a mixed-initiative dialog.

\begin{figure}
\center
{\footnotesize
\begin {tabular}{|p{6.3in}|}\hline
\begin{eqnarray}
\frac{PE|C|A}{(x:PE|C|A|\theta)} = x \label{rule1}\\
\langle \frac{PE|C|A}{(a:T)} \bullet a  = \theta \rangle \label{rule2}\\
\langle \frac{PE}{(x:PE|C|A|T)^{*}(a:T)(y:PE|C|A|T)^{*}} \bullet a \rangle = \frac{PE}{xy} \label{rule3}\\
\langle \frac{C}{(a:T)(x:PE|C|A|T)^{*}} \bullet a \rangle  =
\frac{C}{x}
\label{rule4}\\
\langle \frac{A}{(x:PE|C|A|T)^{*}(a:T)(y:PE|C|A|T)^{*}} \bullet a
\rangle =
\theta \label{alternator-token}\\
\langle \frac{PE}{(x \vdash A)^{*}(y \vdash A)(z:PE|C|A|T)^{*}}
\bullet a \rangle = \frac{PE}{x \langle y \bullet a \rangle z} &
\langle y \bullet a \rangle \neq y
\label{pe-over-alternator}\\
\langle \frac{PE}{(x:PE|C|A|T)^{*}(y:PE|C|A)(z:PE|C|A|T)^{*}}
\bullet a \rangle  = \frac{C}{ \langle y \bullet a \rangle
\frac{PE}{xz}} & \langle y \bullet a \rangle \neq y
\label{rule5}\\
\langle \frac{C}{(x:PE|C|A)(y:PE|C|A|T)^{*}}\ \bullet a \rangle  =
\frac{C}{ \langle x
\bullet a \rangle y} \label{rule6} & \langle x \bullet a \rangle  \neq x \\
\langle \frac{A}{x_{1}^{*}(y_{1}:PE|C|A)x_{2}^{*}\ldots
x_{n}^{*}(y_{n}:PE|C|A)x_{n+1}^{*}} \bullet a \rangle   =
\frac{A}{ \langle y_{1}\ \bullet a \rangle  \langle y_{2} \bullet
a \rangle \ldots \langle y_{n} \bullet a \rangle}
& \langle y_{i} \bullet a \rangle \neq y_{i} \wedge \langle
x_{i} \bullet a \rangle = x_{i}
\label{alternator-subdialog}\\
\langle \frac{PE}{(x:PE|C|A|T)^{*}} \bullet * \rangle  =
\frac{PE}{x}
\label{rule7}\\
\langle \frac{C}{(x:PE|C|A|T)^{*}} \bullet * \rangle  =
\frac{C}{x}
\label{rule8}\\
\langle \frac{A}{(x:PE|C|A|T)^{*}} \bullet * \rangle  =
\frac{A}{x} \label{alternator-error}
\end{eqnarray}
\\\hline
\end{tabular}
}
\caption{Reduction rules for simplifying dialog specifications.}
\label{reduction-rules}
\vspace{-0.2in}
\end{figure}

\subsubsection*{Transformation Rules} 
With this notation in place, it is now possible to present the
rules that govern the behavior of the stagers when
processing user input. Notice that this is rather
tricky as it might require a global restructuring of the representation.
Consider a breakfast dialog given by:
$${\frac{PE}{\frac{C}{e_1\:e_2}\frac{C}{c_1\:c_2}\frac{C}{b_1\:b_2}}}$$
where $e_1, e_2$ are egg specification aspects, $c_1, c_2$ support coffee
specification, and $b_1, b_2$ specify a bakery item. The top-level PE signifies
that the three subdialogs can be entered in any order; the C's denote
that once entered each subdialog involves a second clarification
aspect. After the user has specified his eggs ($e_1$), a clarification of
`how do you like your eggs?' ($e_2$) might be needed. Similarly, when the user
is talking about coffee ($c_1$),  a clarification of `do you take cream and sugar?' ($c_2$)
might 
be required, and so on. Now, assume we stage this dialog using the sequence:
$\prec\!c_1\:e_1\:c_2\:\cdots\!\succ$;
the occurrence of $e_1$ is invalid according to the dialog specification above,
but we will not know that such an input is arriving at the time we are processing
$c_1$. So in response to the input $c_1$, the dialog must be 
restructured as follows:
$${\frac{PE}{\frac{C}{e_1\:e_2}\frac{C}{c_1\:c_2}\frac{C}{b_1\:b_2}}} \bullet c_1
\rightarrow
{\frac{C}{\frac{C}{c_2}\frac{PE}{\frac{C}{e_1\:e_2}\frac{C}{b_1\:b_2}}}}$$
By replacing the top-level PE stager with a C, it
becomes clear that the only legal input
now possible must have $c_2$. Once the coffee subdialog is completed, the top-level
stager will revert back to a PE. Such dialog restructurings are necessary
if we are to remain faithful to the original specification.  We formally
capture such restructurings by using transformation rules
to describe what happens to a 
\{dialog script,
stager\} pair when a given input is received.  

In order to facilitate the description of these rules, the following
notation is used:
\[\frac{PE}{(a:T)}\]
This expression represents a dialog that consists of the dialog
script \(a\) and is being staged with a partial evaluator.
\((a:T)\) indicates that \(a\) is a simple prompt,
not a subdialog.  The expression can be read, `This dialog is
being staged using a \(PE\) and consists of the prompt \(a\), of
type \(T\).' Similarly,
\[\frac{C}{(y:PE|C|A|T)}\]
means that the dialog is staged using a curryer.
\((y:PE|C|A|T)\) indicates that the dialog
script is either a single prompt (T), or a subdialog itself being staged using a partial
evaluator, a currier, or an alternator. The
expression  \(\langle x \bullet a \rangle \), as above, denotes the
transformation of \(x\) when given the input \(a\), as mandated by
the transformation rules.  Finally \(x \vdash A\) indicates that
the subdialog \(x\) contains an alternator somewhere inside it (i.e., at a level
below the top-level stager). 

These rules are shown in
Figure~\ref{reduction-rules}, numbered for convenience.
Rule~\ref{rule1} represents a simplification rule that should be
repeatedly applied to the result of every reduction until no
further simplification is possible. The rest of the rules
represent the reductions that should take place when some input is
given. These rules are listed in order of precedence, so that
the first applicable one fires.

Rule 2 generates the empty dialog ($\theta$) when the given input
matches the only remaining prompt, irrespective of the stager. Rules 3--5 test
if the given input is legal under the current top-level stager and generate
a pruned dialog. Rule 6 is specifically designed for dialogs involving
non-categorical variables, and is discussed further below. Rule 7 handles
the type of transformation such as in the breakfast dialog below. Similar
transformations for a top-level C and A stagers are given by rules 8 and 9.
Rule~\ref{rule7} will fit any input to a dialog script that is
being staged with a partial evaluator; respectively for rules~\ref{rule8}
(currier) and rule~\ref{alternator-error} (alternator).
These three rules represent the
transformation that occurs when no input can be filled, with the
transformation simply generating the original \{dialog script,
interaction stager\} pair. The one other aspect of these rules
that needs to be mentioned is that the result of concatenating
\(\theta\) to any \((x:PE|C|T)\) is \(x\), which will occur when a
dialog or subdialog is completed.

The function of rule~\ref{pe-over-alternator} is worth explaining
in more depth.  In general, once a subdialog is entered, 
the user should have to
complete the subdialog before continuing with the rest of the
dialog.  This requirement is enforced by rule~\ref{rule5}.  When
using the alternator however, this restriction is not always
desirable.  In some situations, the dialog designer may want to
only allow the user to pick one choice from a list of possible
inputs to a dialog, but may still want the top-level dialog to continue
without entering the subdialog. This is especially useful in
non-categorical websites. Rule~\ref{pe-over-alternator}
enables this behavior by checking for a list of alternators on the far
left of the list of subdialogs (when using a PE stager), and uses this
information as a cue to not enter these subdialogs.
This allows the dialog designer to choose either behavior as he 
desires.

\begin{figure}
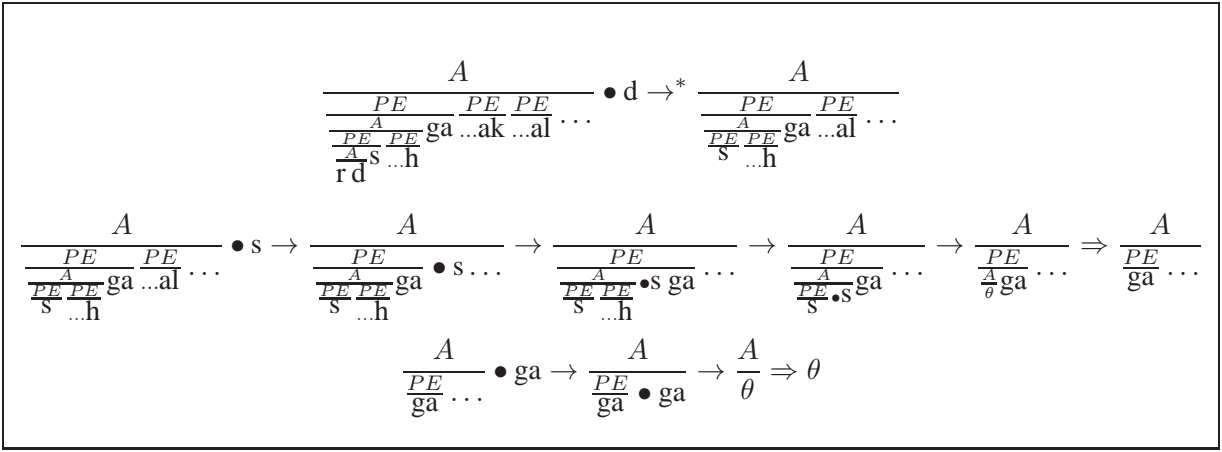

\centering
\begin {tabular}{|p{6.2in}|}\hline
\[
    \frac{A}
    {\frac{PE} {
        \frac{A}
        {\frac{PE} {
            \frac{A} {
                \textrm{r}\;
                \textrm{d}}
            \textrm{s}}
        \frac{PE} {\ldots \textrm{h}}
        }
        \textrm{ga}}
    \frac{PE}{\ldots \textrm{ak}}
    \frac{PE}{\ldots \textrm{al}}
    \ldots
    } \bullet \textrm{d} \rightarrow^{*}
    \frac{A}
    {\frac{PE} {
        \frac{A}
        {\frac{PE} {
            \textrm{s}}
        \frac{PE} {\ldots \textrm{h}}
        }
        \textrm{ga}}
    \frac{PE}{\ldots \textrm{al}}
    \ldots
    }
\]
\[
     \frac{A}
    {\frac{PE} {
        \frac{A}
        {\frac{PE} {
            \textrm{s}}
        \frac{PE} {\ldots \textrm{h}}
        }
        \textrm{ga}}
    \frac{PE}{\ldots \textrm{al}}
    \ldots
    } \bullet \textrm{s} \rightarrow
    \frac{A}
    {\frac{PE} {
        \frac{A}
        {\frac{PE} {
            \textrm{s}}
        \frac{PE} {\ldots \textrm{h}}
        }
        \textrm{ga}} \bullet \textrm{s}
    \ldots
    } \rightarrow
    \frac{A}
    {\frac{PE} {
        \frac{A}
        {\frac{PE} {
            \textrm{s}}
        \frac{PE} {\ldots \textrm{h}}
        } \bullet \textrm{s} \;
        \textrm{ga}}
    \ldots
    } \rightarrow
    \frac{A}
    {\frac{PE} {
        \frac{A}
        {\frac{PE} {
            \textrm{s}}
        \bullet \textrm{s}}
        \textrm{ga}}
    \ldots
    } \rightarrow
    \frac{A}
    {\frac{PE} {
        \frac{A}
        {\theta}
        \textrm{ga}}
    \ldots } \Rightarrow
    \frac{A}
    {\frac{PE} {
        \textrm{ga}}
    \ldots}
\]
\[
    \frac{A}
    {\frac{PE} {
        \textrm{ga}}
    \ldots} \bullet \textrm{ga} \rightarrow
    \frac{A}
    {\frac{PE} {
        \textrm{ga}} \bullet \textrm{ga}} \rightarrow
    \frac{A}
    {\theta} \Rightarrow
    \theta
\]
\\\hline

\end{tabular}
\caption{Staging the interaction of Fig.~\ref{votesmart-oot} using
transformation rules.}
\label{reduction-examples}
\end{figure}
\begin{figure}[t]
\centering
{\small
{\tt
\begin{tabular}{|l|l|l|} \hline
(1)~~~read (r, h); & read (r, h); & read (r, h);\\
(2)~~~cArea = $\pi$*r$^{2}$; & cArea = $\pi$*r$^{2}$; &\\
(3)~~~sArea = 2*cArea+2*r*$\pi$*h; & & sArea = 2*cArea+2*r*$\pi$*h;\\
(4)~~~vol = cArea*h; & vol = cArea*h; & vol = cArea*h;\\
(5)~~~print (sArea); & & print (sArea);\\
(6)~~~print (vol); & print (vol); & print (vol);\\
\hline
\end{tabular}}}
\caption{Illustration of program slicing.
(left)~A program which takes the radius and height of a cylinder
as input and computes and prints its surface area and volume.
(center)~A static backward slice~(of~left) w.r.t.~(6,~{\tt vol}).
(right)~A static forward slice~(of~left)~w.r.t.~(1,~{\tt h})
(variable key: r~=~radius;~h~=~height;~cArea~=~circle area;~sArea~=~surface
area;~vol~=~volume).}
\label{slicing}
\end{figure}

We now demonstrate the staging of the interaction in
Fig.~\ref{votesmart-oot} using the transformation rules (see Fig.~\ref{reduction-examples}).
The beginning dialog is similar to the modeling
of the website shown earlier, but to trigger the application of
rule~\ref{pe-over-alternator} the subdialogs are written in the
form: $\frac{PE}{<\!\mathrm{page}\!> <\!\mathrm{link text}\!>}$
with the link text appearing to the {\it right} of the page description (the
meaning is still preserved since the stager is a PE). These substructures
thus represent a link label and the page that the label leads to (when
dealing with a website that is not in tree
form, the page could be a reference to a page defined elsewhere). Recall
that this `page' is typically a subtree in the original website.

In order to preserve space, subdialogs that will not be entered are not
completely shown.  In some cases though, they are assumed to
contain certain attributes that prevent them from being removed
from the dialog. For example, it is assumed the Alabama (`al') has
some politicians who are Democrats (`d'), and thus remains in the
dialog after the utterance `d.'  On the other hand, Alaska (`ak'),
does not, so it is simplified out of the dialog when the user says
`d'.  While Alabama does have Democratic politicians, none of them
are senators, and is thus simplified out of the dialog
after the user specifies Senator (`s').

After every input, the appropriate transformation rules are
applied, and the resulting \{dialog script, interaction stager\}
pair is simplified.  The simplified dialog is then used as the
model to accept the next piece of input.  This process is carried
out until the dialog is reduced to \(\theta\), indicating that the
dialog has completed. Each $\rightarrow$ in Fig.~\ref{reduction-examples}
describes a transformation based on user input, and each $\Rightarrow$
describes a simplification of the dialog structure.

For the first interaction, all of the
transformations are not shown for readability.  The second
interaction is shown in more detail. First
rule~\ref{alternator-subdialog} is applied, which removes Alabama
from the dialog, since it is no longer applicable. This is
followed by rule~\ref{pe-over-alternator} to begin simplifying the
Branches page, followed by rule~\ref{alternator-subdialog}, which
removes House (`h') from the dialog.  The next transformation is
based on rule~\ref{rule2}, which removes the need to say Senate
from the dialog, since it has already been said.  The resulting
dialog is then simplified as per rule~\ref{rule1} (denoted by
\(\Rightarrow\)), to yield the new dialog.

\subsubsection*{Building a Robust Transformation Engine}
We have presented a formal theory for reasoning about hierarchical
staging notation and for simplifying dialogs; to target this theory
for real-time interaction management in websites, we represent interactions
with websites as XML documents and implement the transformations using
XSLT technology~\cite{XSLTbook}. The XML documents summarize the
hyperlink structure, the hierarchical staging notation, and indirectly
the vocabulary comprising of legal patial input.
While XML documents obey a tree-structured model, notice that we can use
id's and refid's to factor crosslinks, subdialogs in other sites,
and hence effectively model DAGs as well.

Recall that the stagers
serve a dual role in our framework: they enforce acceptability
criteria on user input (i.e., by distinguishing valid from invalid inputs) and they
also capture the underlying program transformation that must be performed, for valid input. 
Therefore, in implementing the framework, we first verify
user input for legality (e.g.,
when the user says `Democrat' out-of-turn, we consult the dialog specification to determine
if this is acceptable) and then perform the desired transformation to accommodate this
partial input. Interestingly, all of 
I, PE, C, and A staging functionality can be implemented in terms
of a more general program transformation called {\it slicing}.

Program slicing~\cite{programSlicing}
is a technique used to extract statements,
which {\it may} affect or be affected by the values of variables of interest
computed at some point of interest, from a program.
A slice of a program is taken w.r.t. a~(a point of
interest, a variable of interest) pair,~referred to as
the {\it slicing~criterion}. The point of interest is specified with a
line number from the program.
The resulting slice consists of all program statements which may affect or be
affected by the value of the variable at the specified point.

Fig.~\ref{slicing} illustrates simple program slicing.  Slices such as that
shown in Fig.~\ref{slicing}~(center), are called~`backward
slices'~\cite{programSlicing}. 
The slice is {\it backward} since this is the direction in which dependencies
are followed to their sources in the program.  Contrast this with a {\it
forward slice}~\cite{InterproceduralSlicing}~which consists of the program
statements affected by the value of a particular variable at a particular
statement~(see Fig.~\ref{slicing} --- right).  Backward slices contain data and
control predecessors, while forward slices consist of data and control
successors.  For an introduction to program slicing
and applications, we refer the interested reader to~\cite{programSlicing}.

The transformation of websites, for I, C, PE, and A stagers, can be modeled
as a forward slice followed by a backward slice (w.r.t. corresponding program variables).
Intuitively, given valid input, a forward slice is performed w.r.t. the corresponding
program variable to determine the terminal webpages that are reachable from that
point. These webpages are collected and back-propagated via backward slicing, so
that only those paths that reach these pages are retained. Notice that these two
operations implicitly capture exclusions among program variables; e.g., when the user
says `Democrat' the slices will remove any program segments that involve Republicans. 
XSLT's support for pattern-oriented programming is particularly
advantageous here since both forward and backward slicing
can be captured in the form of ancestor or descendant axis types in location paths.
Such a combination of forward and backward slicing is closely related to other 
operators for pruning information hierarchies, namely Sacco's 
zoom operator~\cite{DynamicTax}.

\vspace{-0.16in}
\subsection{Interaction Interfaces}
To exercise our staging transformation framework, we developed two different
input interfaces. The first ({\it Extempore}), leverages
users' familiarity with toolbar interfaces, and provides a way to supply
out-of-turn textual input. The second ({\it SALTII}, for SALT Interaction Interface,
pronounced `salty') utilizes a rapidly emerging technology for integrating
speech into web browsing. Both interfaces employ a common Javascript
toolkit that handles communication with the interaction manager (see next
section), and which is designed to reduce the development cost for future 
interaction interfaces (e.g., PDAs and 3G phones).
\begin{figure}
\centering
\vspace{-0.2in}
\begin{tabular}{c}
\includegraphics[width=13cm]{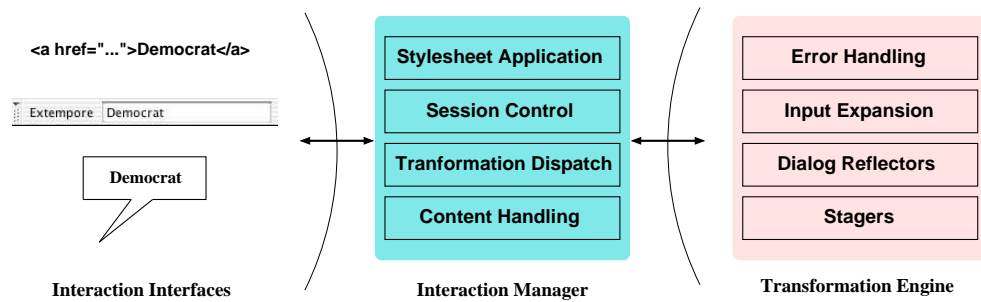}
\end{tabular}
\centering
\caption{Multimodal web interaction framework architecture, depicting the
central role played by the interaction manager.}
\label{im-diagram}
\end{figure}

\vspace{-0.05in}
\subsubsection*{Extempore}
The Extempore toolbar was developed using the XML
User interface Language (XUL) and JavaScript for the cross-platform Mozilla web browser. 
It was designed to be non-invasive and to become active only 
when the user is visiting a website capable of out-of-turn interaction. By
displaying a lightweight, text-based interface, Extempore leverages users'
prior knowledge to provide a familiar and easy method of interaction. 
We will see an illustration of Extempore in a new case study depicted in
Fig.~\ref{odp-demo}.
It is important to note that Extempore is embedded in the web browser,
and not the site's webpages. It also is not a site-specific search tool
that returns a flat list of results~(akin to the Google toolbar).

\vspace{-0.05in}
\subsubsection*{SALTII}
The SALTII interface is built using the SALT 
XML-based markup language, allowing one to embed speech tags in HTML to
realize webpages capable of speech input and output.
The current SALTII implementation
requires the SALT voice recognition plugin for Microsoft Windows Internet 
Explorer 6.0. Using this interface, users could potentially carry out an
entire dialog with speech alone, using speech for not only out-of-turn interaction,
but in-turn as well.
This interaction interface is patterned after speech recognition technologies
such as VoiceXML. 

\vspace{-0.05in}
\subsubsection*{Future Interfaces}
The JavaScript toolkit for the above interfaces was designed with future 
development in mind and factors the entire system to roughly five simple 
functions. Since it is developed using the vcXML-RPC JavaScript library, the
functionality of the interaction manager can be remotely accessed, enabling
shared context scenarios~\cite{PUI}. This toolkit has been made available 
at the project's website (http://pipe.cs.vt.edu)
for developers of other interaction interfaces.

\subsection{Interaction Manager}
The interaction manager primarily co-ordinates communication between the 
transformation engine and the interaction interfaces. Recall that the
staging transformations framework treats in-turn inputs no differently from 
out-of-turn inputs, so it is desirable that the interaction manager also
preserve this uniformity. We first outline the overall process by which 
interaction is established and managed, followed by descriptions 
of the four constitutent subsystems (see Fig.~\ref{im-diagram}).

\vspace{-0.05in}
\subsubsection*{Preparing for Out-of-Turn Interaction}
To situate the interaction manager as a dialog facilitator of
both in-turn and out-of-turn inputs, we have investigated a variety of mechanisms, 
ranging from those that involve the full participation of the website, to 
proxy-based bypass schemes. The former requires a DNS re-direction so that HTTP 
GET requests are forwarded to the interaction manager (notice that out-of-turn inputs 
are received directly from the interaction interfaces). This solution also has the 
attractive property that mixed-initiative interaction can be enabled at as fine or 
coarse a level of granularity as desired (e.g., it can be enabled for
only certain subtrees). The proxy-based approach is a less configurable solution
and must be targeted very carefully, to avoid loss of functionality. We adopt
the former approach in this paper. Once such an initial handshake is established,
the interaction manager is responsible for providing concurrent access to 
the transformation engine, from potentially multiple interaction interfaces. 

The interaction manager, now placed in the loop, evaluates if out-of-turn
interaction is possible, activates the interaction interfaces as appropriate,
and mediates all interactions from this point. Notice that intermediate dialog
states might not correspond to any of the site's existing webpages (especially after
some out-of-turn interaction), so the interaction manager must mediate the dialog
to the fullest. Since we target only hierarchical sites in this paper, the interaction
manager need revert back to the original site only to display the leaf page(s).

\vspace{-0.05in}
\subsubsection*{Content Handling}
Content handling determines the feasibility of out-of-turn
interaction, caches dialog states, and ensures currency of site
representations. It is also responsible for retrieving, caching, and updating
content from websites.

To determine the feasibility
of out-of-turn interaction, the content handler
uses a simple HTTP GET request for a well-formed XML document named
`content.xml' located at the website's root directory (e.g., this file 
for the website http://www.\hskip0ex smallbox.org would be at the location 
http://www.smallbox.org/\hskip0ex content.xml). This document, if it
exists, is meant to supply the representation of the site, and when annotated with
stager tags, helps initialize the dialog representation. It is then stored
in a local database for fast transformation computations. It also initiates the
activation of the Extempore toolbar or the SALT tags, as appropriate. 
From this point, the content handler is responsible for ensuring
the currency of the representation and re-retrieving the file as appropriate.

Notice that caching is trivially implemented by associating intermediate dialog 
states to content files generated over the course of an interaction. A more
sophisticated solution is to develop a caching policy that exploits the structure
of program transformations. For instance, if a user is requesting a partial
evaluation w.r.t. `Democratic Senators,' but the cache only contains a document
that has been evaluated w.r.t. `Democrat,' we can partially evaluate this document
internally w.r.t. the remaining input (namely, `Senate'), thus removing the need to 
partially evaluate from the root document. While reducing storage complexity
this approach also 
creates interesting design tradeoffs (including concerns about session and user
security).

\vspace{-0.05in}
\subsubsection*{Transformation Dispatch}
The transformation dispatch is responsible for handling communication 
with the transformation engine. It handles connecting to the transformation 
engine as well as notifying the interaction interfaces if such a connection 
cannot be made (When the transformation engine receives partial input,
recall that it does not know, and need not know, whether the partial input is a result of
browsing or of supplying some information out-of-turn).
Finally, transformation dispatch supports the marshalling and un-marshalling of
transformation requests into messages, as well as the transmission and 
reception of those messages.

\vspace{-0.05in}
\subsubsection*{Session Control}
Session control has an interesting responsibility that differs from 
most web systems' concept of session management. 
Notice that our notion of `state' in a dialog is just the staging representation,
since it succinctly summarizes all remaining dialog options. Furthermore, the
transformation engine does not explicitly manipulate state and is hence, a purely
functional\footnote{We use the word `functional' in a programming languages
connotation (e.g., Haskell).} entity. 
Thus the goal of session control is merely to distinguish
one user's interaction from another. 
Due to our requirement for handling in-turn and out-of-turn inputs uniformly,
session tokens (we use a ten decimal digit identifier) are required to be
kept in two different places, the interaction interface and the browser 
itself. This two-headed
session format negates the application of most modern session 
management packages, which are primarily concerned with tracking browsing
interactions. A session manager was specifically designed to 
handle this issue as well as to handle the normal
session management issues (e.g., back button browsing and threaded browsing).

\vspace{-0.05in}
\subsubsection*{Stylesheet Application}
Stylesheet application is responsible for transforming the 
information returned from the transformation engine into the site's native 
presentation format. In addition, it must introduce suitable grammar tags
into the HTML page (for the voice interface) by analyzing the remaining dialog options.
Currently we support [X]HTML, WML, SALT, SVG, or 
any XML-based presentation 
format, and this is determined by the interaction interface making the request. 
Default stylesheets for these formats are made publicly available 
from the project website. 

\subsection{Miscellaneous Design Decisions}
\vspace{-0.05in}
\subsubsection*{Input Validation}
Input validation for the Extempore toolbar interface is performed directly
by the reduction rules of Fig.~\ref{reduction-rules} whereas input validation
for the voice interface is trapped on the client side by suitable
generation of an SRGS grammar (at the Stylesheet Applicator).

\subsubsection*{Orienting Users}
To better orient the user in their interactions, we implemented an `Input So Far:'
feedback label (in the browser status bar) in both implementations that summarizes the 
partial input supplied thus far (e.g., see Fig.~\ref{votesmart-oot}). We also support 
meta-dialog enquiries (e.g., `What may I say?') via dialog reflectors. This is activated
through a `?' button in Extempore and by a spoken query in SALTII, and involves
traversing the current representation for determining legally specifiable inputs.

\subsection{Implementation Details}
The elegance of our implementation is reflected in the minimal codebase required.
The transformation engine is built using C and the libxml, libxslt libraries, and
is under 500 lines of code. This invokes a 120-line transformation template for
out-of-turn interaction, with 50 lines for handling input expansion and dialog
reflection capabilities. The transformation engine is wrapped using SOAP (Simple 
Object Access Protocol), effectively making it a web service~\cite{webservices2,webservices1} from the 
perspective of the interaction manager.
External communication thus happens through SOAP messages. 
Extempore is implemented in 50 lines of XUL and
SALTII only requires lines of code proportional to the size of the underlying
grammar. The JavaScript toolkit supporting new interfaces is about 300 lines of code.
The interaction manager was developed using the PHP scripting language, and is
implemented in a total of 375 lines of
code, barring two external open source libraries (the NuSOAP library and the 
vcXML-RPC library, for communication between other layers). 

\begin{figure}
\centering
\vspace{-0.2in}
\begin{tabular}{c}
\includegraphics[width=8.4cm]{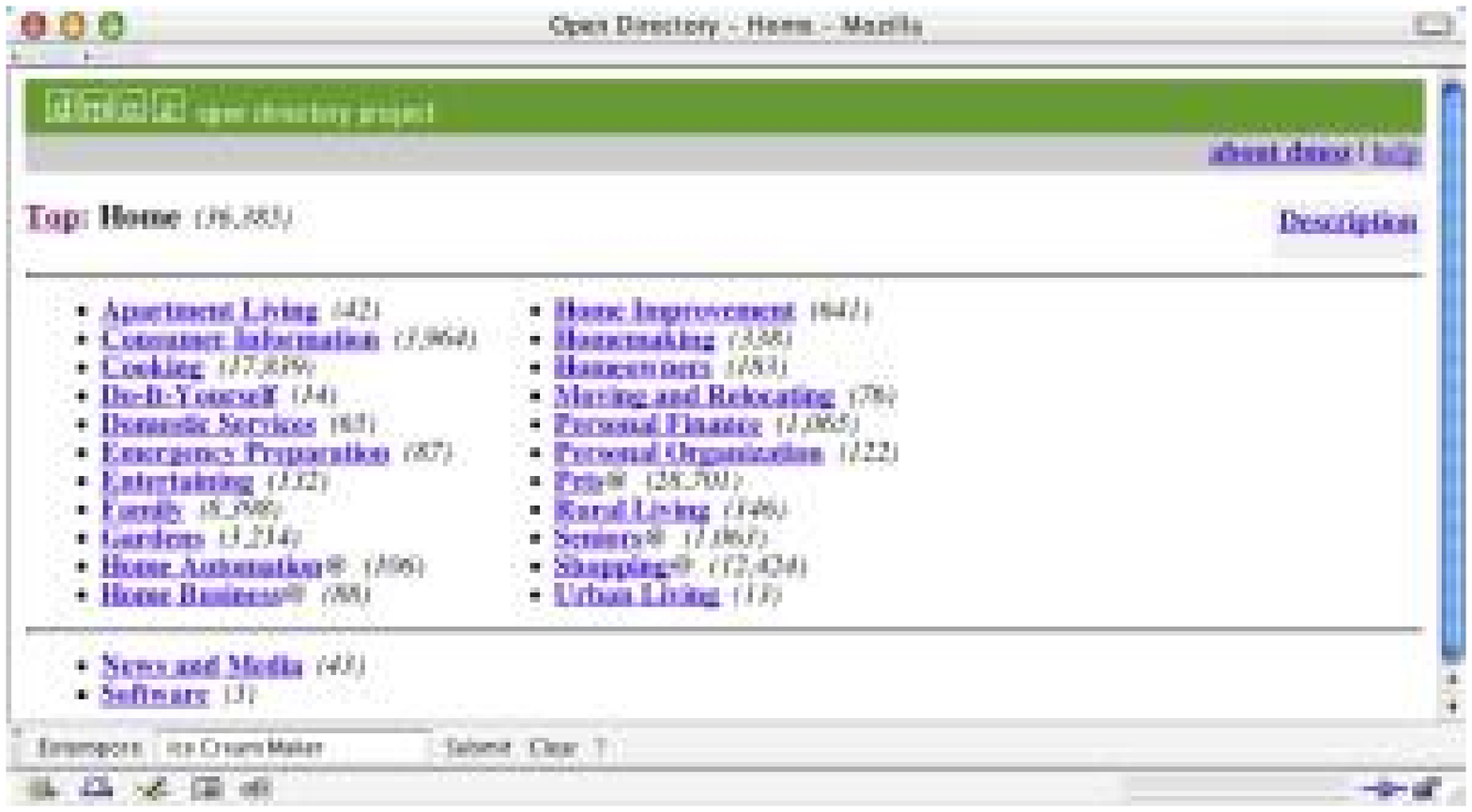}\\
$\mathbf{\Downarrow}$\\
\includegraphics[width=8.4cm]{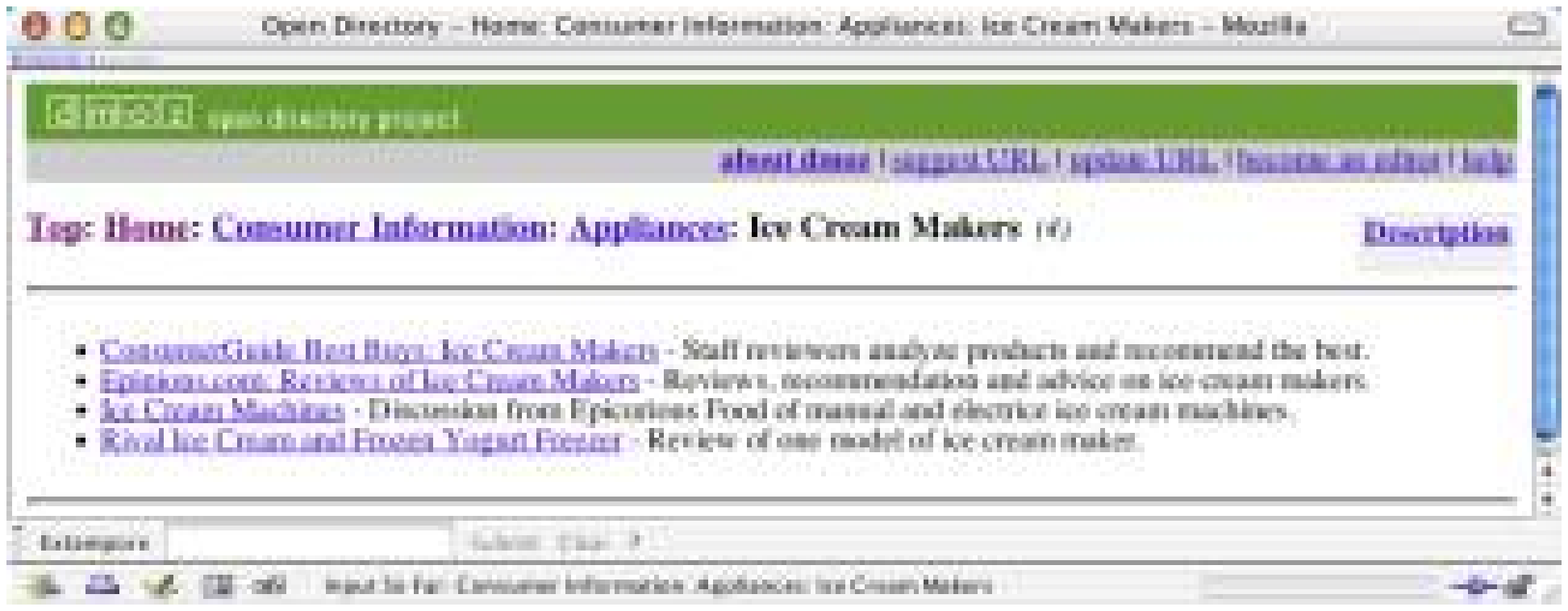}\\
\end{tabular}
\centering
\caption{Out-of-turn interaction with the HOME subtree of the ODP hierarchy. In
the top window, the user enters `Ice Cream Maker' out-of-turn using the Extempore
toolbar, producing the page shown below. Notice
that two functional dependencies are automatically triggerred: 
\{Ice Cream Makers\} $\rightarrow$ \{Appliances, Consumer Information\}.}
\label{odp-demo}
\end{figure}

\vspace{-0.1in}
\section{Application Case Studies}
Besides the Vote Smart study used as our running example, we implemented
multimodal web interfaces for selected subtrees of the ODP hierarchy (dmoz.org) and
the CITIDEL repository of research papers (citidel.org). Due to space considerations,
we only discuss the Vote Smart and ODP applications here. 

Figs.~\ref{votesmart-it} and~\ref{votesmart-oot} have already depicted interactions
with the multimodal web interface to Vote Smart. 
Here, we actually employed a four-level modeling 
(state, branch, party, district/seat) although we have described only the first three 
levels (state, branch, party) thus far in this paper, for ease of description.

Fig.~\ref{odp-demo} depicts a multimodal web interaction with
the ODP HOME subtree, this time using Extempore. ODP is a human-compiled directory
of the web and is a constantly evolving categorization of websites. A significant
proportion of the links in ODP are actually symbolic links, where one subtree points to
another, under a different classification. For instance, `Recreation: Martial Arts'
actually points to the `Martial Arts' subtree physically located under `Sports.'
It can be argued that the presence of symbolic links is actually foretelling of the
mental mismatch anticipated by directory compilers, and hence {\it in-turn} mechanisms
have been hardwired to better orient users. The use of out-of-turn interaction,
in effect, obviates the need for symbolic links by increasing the flexibility of communicating
partial input.

\subsubsection*{Input Expansion}
In building both these applications, we have realized the importance of
input expansion strategies to
capture semantic relatedness among dialog
options. An example of a semantic constraint is when the user says 
`Senior seat' out-of-turn, we can also infer the choice of `Senate' (as opposed to 
`House'). This is because the functional dependency \{Senior seat\} $\rightarrow$ 
\{Senate\} holds in the underlying domain.
Such dependencies can be automatically inferred by simple association
mining methods~\cite{autoPersWebMining}; we identified 125 of them in the 
Vote Smart site and more than 3,600 in just the ODP's Home subtree. 
An elegant by-product of such automatic
input expansion is that out-of-turn interaction can provide rapid shortcuts to desired
leaves, even at the root level. For instance, Washington, D.C. has only one 
representative and no senators, and hence the following dependency holds:
\{Washington, D.C.\} $\rightarrow$ \{House, Democrat, District at Large\}.
Therefore, saying/clicking `Washington, D.C.' at the top-level page uniquely identifies
one congressperson and transports the user directly to her webpage. 
Mining functional dependencies for use in out-of-turn interaction
is an interesting topic by itself, and one that we intend to explore fully in 
a future paper. 

\subsubsection*{Collapsing Transformations}
Another practical consideration exists when only one leaf page (e.g., congressperson)
remains as 
the result of a transformation. For instance, in the current political
landscape, the Democratic senator from Minnesota is occupying the senior seat, 
therefore saying `Minnesota Democrat Senator' uniquely identifies a politician.
In such a case, we collapse the remaining
series of hyperlinks (involving seat) and return the webpage of the official directly 
to the user~(relieving
her from having to click through these links).  

Notice that no information
is lost as a result of either automatic input expansion or collapsing as a 
leaf webpage contains the values for the facets under which it is classified.

\vspace{-0.05in}
\section{Evaluation}
There are many ways to evaluate the framework presented here. For instance, we can
study user experiences with the deployed applications, characterize the framework
by its support for modeling, and also investigate the ease of implementing new 
applications within the framework.

User experiences with out-of-turn interaction are described in~\cite{ExtemporeCHI}; 25 users
were given information-finding tasks about U.S. politicians and were free to
use either in-turn or mixed-initiative interactions to complete
these tasks. Some of these
tasks were {\it non-oriented} (meaning they could be performed with browsing alone, if
desired) and some were {\it out-of-turn-oriented} (meaning
they would be cumbersome to perform via plain browsing). We found that 100\% of the users 
utilized the out-of-turn interfaces when presented with an out-of-turn-oriented task.
Since the task type was not disclosed {\it a priori}, this result
demonstrates that users are adept at discerning when out-of-turn interaction is 
desirable. Extempore and SALTII interfaces were utilized equally effectively.

From an information-seeking standpoint, it is easy to see that our use of out-of-turn
interaction dramatically
increases the number of ways to reach a given webpage. For example, in Vote Smart,
the original 540 browsing sequences are now a small subset of 
12,960 realizable sequences (540$\times$~4!), where the length of 
each interaction sequence is constant (this is not always the case; e.g., ODP).
This is a 2,300\% increase in the number of sequences
supported!  From a representational perspective, such increase comes through
{\it without} modeling any more than the original 540 sequences. This is in
contrast to faceted browsing~\cite{Flamenco} which would cause an exponential
blowup in site structure. In fact, we do not even model the original 540 distinctly as
our nesting of dialog choices factors the representation. Similarly, in ODP
we observe a 44,400\% increase in the number of sequences supported in relation to
the 1,411 original browsing sequences.

New applications, especially hierarchical sites, are easily targeted using the
software framework presented here (see project website for more information). 
Taxonomies (e.g., gams.nist.gov), LDAP directories, database-driven indices (e.g.,
acm.org/dl), and bioinformatics ontologies (e.g., www.gene-ontology.org) are ideal 
for modeling in the staging transformations framework. The property that these information 
sources share is that they all foster (and sometimes require) 
{\it focused} information-seeking behavior.

In contrast, consider a site such as the Internet movie database (www.imdb.com), which
is meant for {\it exploratory} browsing
and uses connections in social networks as a navigational metaphor. Such a non-hierarchical
site is more cumbersome to model using staging transformations. 

Another shortcoming of staging transformations is that the partial input suppliable
by the user is primarily of a {\it declarative} nature, and hence does not adequately
support {\it procedural} tasks. For instance, consider a task such as `Find the 
political party of the senior senator representing the only state which has congresspeople 
from the Independent party.' It involves finding Vermont as the answer to the 
`only state part' (via out-of-turn interaction) and then using it in {\it another} 
interaction to find the political party of the senior senator from that state. When
tested with participants, we found that only 50\% of the users successfully completed
this task~\cite{ExtemporeCHI}, because the rest of them were attempting to continue 
the interaction after answering `Vermont' for the first part of the question. 
To support such prolonged dialogs, staging tranformations must provide {\it
constructive operators}; all staging operators considered in this paper 
are {\it destructive} in that valid partial input causes pruning 
of remaining dialog options.

A final technological limitation pertains to speech interaction in large sites
(e.g., higher subtrees of ODP). In a grammar based approach such as used in SALT, 
large sites can involve a dramatic growth in vocabulary, especially when we use
a PE stager at a high-level of composition. More robust statistical methods or
other dialog abstraction capabilities must be investigated.

\vspace{-0.05in}
\section{Related Research}
Web interaction management emerged as a legitimate area
of research ever since researchers attempted to build stateful
and responsive web applications on top of stateless protocols such as
HTTP~\cite{graham-cacm-interview}. Interaction management research is typically 
concerned with issues such as automated delivery of static as well
as dynamically generated pages~\cite{staticPages}, 
sessioning, accommodating simultaneous users, concurrency control, 
stateful implementation of client-side functionality such as cloning windows and
pursuing back buttons~\cite{continuationsWebServers}, and domain-specific 
language (DSL) support for targeted applications such as form-field
interaction and database-centered services (e.g., {\tt <bigwig>}~\cite{bigwig}
and MAWL~\cite{mawl}).
Interestingly, a significant
body of this research has involved concepts from functional programming
(e.g., continuations, currying)~\cite{automaticallyRestrWPs,uic,continuationsWebServers}.

Our work embraces this tradition and proposes the use of program transformations for
staging web dialogs. It thus casts the problem of dialog control and
management in a purely functional framework, with attendant benefits. The modeling 
of interaction undertaken here is reminiscent of the approach advocated 
by Marchionini for designing information systems~\cite{ISinEE}. It also addresses 
Dumais's vision of a tighter coupling between structure and search in 
information access~\cite{dumais-structure-search}. 

Another pertinent area of related research can be found in the adaptive hypermedia
community~\cite{webCompanions,adaptiveHypermediaSurveyJoP2,adaptiveHypermediaCSUR}. Here, an explicit user model is
built (e.g., from past interactions) which is then used as the driver to support
adaptive presentation and personalized interaction. Our philosophy, on the other hand, 
is that by enriching the expressiveness with which users can supply partial input, 
we can help them achieve their information-seeking goals better. Needless to say, these 
two views can be fruitfully integrated.

The software framework proposed here is complementary to other frameworks
for interaction co-ordination~\cite{CoordMultimodal,dyna-coord,web-middleware},
functional web adaptation~\cite{functional-web-adapt} and re-engineering of 
websites~\cite{automaticallyRestrWPs}. However, the specific setting assumed here (i.e.,
out-of-turn interaction) is different from those considered in these works. Our
framework is closer in motivation to systems like VoiceXML~\cite{VoiceXMLW3C} which
provide support for creating mixed-initiative dialogs.

\vspace{-0.05in}
\section{Discussion}
We have described a software framework for
multimodal out-of-turn interaction, thus laying the foundation for creating
mixed-initiative dialogs with websites. Our usage of out-of-turn interaction is optional, 
unintrusive, and can be integrated into browsing experiences at the user's
discretion. It also promotes a novel interpretation of multimodal paradigms.
For the designer, the framework simplifies the process of integrating
in-turn and out-of-turn interaction using a uniform handling of both dialog
specification and implementation. Minimal modeling is required to re-engineer
existing sites. 

The multimodal web view realized in this paper also 
extends the idea of web access via voice~\cite{WebViews,AVoN} and could be usefully
applied in a variety of mobile browsing contexts 
(e.g., see~\cite{know-encap-pervas,whole-parts,pers-pocket-dirs}).
The framework was developed with future web access paradigms in mind, beyond
cell phones and PDAs. The factoring of the system architecture into 
three components means that content providers and developers
need only concern themselves with the interfaces and modalities they wish to support.
An additional possibility we are exploring is the idea of supporting
interjection-style out-of-turn interaction, wherein the browser can dynamically
update content {\it while} the user is supplying out-of-turn input.
This feature is currently projected to utilize the SSU~\cite{ssu} software framework to 
provide real-time feedback to users, and alert them if something is amiss.
The current model of interaction management uses a coarser level of dialog unit
at which feedback is provided. Exploring these issues will undoubtedly open up
new research directions and additional applications for multimodal web interfaces.

\vspace{-0.05in}
\section{Acknowledgements}
The authors thank Manuel A. P\'{e}rez-Qui\~{n}ones, Robert Capra (Virginia Tech), 
Atul Shenoy (Microsoft),
and Christian Queinnec 
(Universit\'{e} Paris) for helpful discussions and comments. 
This work
is supported in part by US National Science Foundation grant IIS-0136182.
\bibliography{www2004}
\bibliographystyle{plain}
\end{document}

%% file: psfig-dvips.tex
\ifx\undefined\psfig\else \fi

\edef\psfigRestoreAt{\catcode`@=\number\catcode`@\relax}
\catcode`\@=11\relax
\newwrite\@unused
\def\typeout#1{{\let\protect\string\immediate\write\@unused{#1}}}
\def\figurepath{./}

\def\@nnil{\@nil}
\def\@empty{}
\def\@psdonoop#1\@@#2#3{}
\def\@psdo#1:=#2\do#3{\edef\@psdotmp{#2}\ifx\@psdotmp\@empty \else
    \expandafter\@psdoloop#2,\@nil,\@nil\@@#1{#3}\fi}
\def\@psdoloop#1,#2,#3\@@#4#5{\def#4{#1}\ifx #4\@nnil \else
       #5\def#4{#2}\ifx #4\@nnil \else#5\@ipsdoloop #3\@@#4{#5}\fi\fi}
\def\@ipsdoloop#1,#2\@@#3#4{\def#3{#1}\ifx #3\@nnil 
       \let\@nextwhile=\@psdonoop \else
      #4\relax\let\@nextwhile=\@ipsdoloop\fi\@nextwhile#2\@@#3{#4}}
\def\@tpsdo#1:=#2\do#3{\xdef\@psdotmp{#2}\ifx\@psdotmp\@empty \else
    \@tpsdoloop#2\@nil\@nil\@@#1{#3}\fi}
\def\@tpsdoloop#1#2\@@#3#4{\def#3{#1}\ifx #3\@nnil 
       \let\@nextwhile=\@psdonoop \else
      #4\relax\let\@nextwhile=\@tpsdoloop\fi\@nextwhile#2\@@#3{#4}}
\newread\ps@stream
\newif\ifnot@eof       % continue looking for the bounding box?
\newif\if@noisy        % report what you're making?
\newif\if@atend        % %%BoundingBox: has (at end) specification
\newif\if@psfile       % does this look like a PostScript file?
{\catcode`\%=12\global\gdef\epsf@start{%!}}
\def\epsf@PS{PS}
\def\epsf@getbb#1{%
\openin\ps@stream=#1
\ifeof\ps@stream\typeout{Error, File #1 not found}\else
   {\not@eoftrue \chardef\other=12
    \def\do##1{\catcode`##1=\other}\dospecials \catcode`\ =10
    \loop
       \if@psfile
	  \read\ps@stream to \epsf@fileline
       \else{
	  \obeyspaces
          \read\ps@stream to \epsf@tmp\global\let\epsf@fileline\epsf@tmp}
       \fi
       \ifeof\ps@stream\not@eoffalse\else
       \if@psfile\else
       \expandafter\epsf@test\epsf@fileline:. \\%
       \fi
          \expandafter\epsf@aux\epsf@fileline:. \\%
       \fi
   \ifnot@eof\repeat
   }\closein\ps@stream\fi}%
\long\def\epsf@test#1#2#3:#4\\{\def\epsf@testit{#1#2}
			\ifx\epsf@testit\epsf@start\else
\typeout{Warning! File does not start with `\epsf@start'.  It may not be a PostScript file.}
			\fi
			\@psfiletrue} % don't test after 1st line
{\catcode`\%=12\global\let\epsf@percent=%\global\def\epsf@bblit{%BoundingBox}}
\long\def\epsf@aux#1#2:#3\\{\ifx#1\epsf@percent
   \def\epsf@testit{#2}\ifx\epsf@testit\epsf@bblit
	\@atendfalse
        \epsf@atend #3 . \\%
	\if@atend	
	   \if@verbose{
		\typeout{psfig: found `(atend)'; continuing search}
	   }\fi
        \else
        \epsf@grab #3 . . . \\%
        \not@eoffalse
        \global\no@bbfalse
        \fi
   \fi\fi}%
\def\epsf@grab #1 #2 #3 #4 #5\\{%
   \global\def\epsf@llx{#1}\ifx\epsf@llx\empty
      \epsf@grab #2 #3 #4 #5 .\\\else
   \global\def\epsf@lly{#2}%
   \global\def\epsf@urx{#3}\global\def\epsf@ury{#4}\fi}%
\def\epsf@atendlit{(atend)} 
\def\epsf@atend #1 #2 #3\\{%
   \def\epsf@tmp{#1}\ifx\epsf@tmp\empty
      \epsf@atend #2 #3 .\\\else
   \ifx\epsf@tmp\epsf@atendlit\@atendtrue\fi\fi}

\chardef\letter = 11
\chardef\other = 12

\newif \ifdebug %%% turn me on to see TeX hard at work ...
\newif\ifc@mpute %%% don't need to compute some values
\c@mputetrue % but assume that we do

\let\then = \relax
\def\r@dian{pt }
\let\r@dians = \r@dian
\let\dimensionless@nit = \r@dian
\let\dimensionless@nits = \dimensionless@nit
\def\internal@nit{sp }
\let\internal@nits = \internal@nit
\newif\ifstillc@nverging
\def \Mess@ge #1{\ifdebug \then \message {#1} \fi}

{ %%% Things that need abnormal catcodes %%%
	\catcode `\@ = \letter
	\gdef \nodimen {\expandafter \n@dimen \the \dimen}
	\gdef \term #1 #2 #3%
	       {\edef \t@ {\the #1}%%% freeze parameter 1 (count, by value)
		\edef \t@@ {\expandafter \n@dimen \the #2\r@dian}%
		\t@rm {\t@} {\t@@} {#3}%
	       }
	\gdef \t@rm #1 #2 #3%
	       {{%
		\count 0 = 0
		\dimen 0 = 1 \dimensionless@nit
		\dimen 2 = #2\relax
		\Mess@ge {Calculating term #1 of \nodimen 2}%
		\loop
		\ifnum	\count 0 < #1
		\then	\advance \count 0 by 1
			\Mess@ge {Iteration \the \count 0 \space}%
			\Multiply \dimen 0 by {\dimen 2}%
			\Mess@ge {After multiplication, term = \nodimen 0}%
			\Divide \dimen 0 by {\count 0}%
			\Mess@ge {After division, term = \nodimen 0}%
		\repeat
		\Mess@ge {Final value for term #1 of 
				\nodimen 2 \space is \nodimen 0}%
		\xdef \Term {#3 = \nodimen 0 \r@dians}%
		\aftergroup \Term
	       }}
	\catcode `\p = \other
	\catcode `\t = \other
	\gdef \n@dimen #1pt{#1} %%% throw away the ``pt''
}

\def \Divide #1by #2{\divide #1 by #2} %%% just a synonym

\def \Multiply #1by #2%%% allows division of a dimen by a dimen
       {{%%% should really freeze parameter 2 (dimen, passed by value)
	\count 0 = #1\relax
	\count 2 = #2\relax
	\count 4 = 65536
	\Mess@ge {Before scaling, count 0 = \the \count 0 \space and
			count 2 = \the \count 2}%
	\ifnum	\count 0 > 32767 %%% do our best to avoid overflow
	\then	\divide \count 0 by 4
		\divide \count 4 by 4
	\else	\ifnum	\count 0 < -32767
		\then	\divide \count 0 by 4
			\divide \count 4 by 4
		\else
		\fi
	\fi
	\ifnum	\count 2 > 32767 %%% while retaining reasonable accuracy
	\then	\divide \count 2 by 4
		\divide \count 4 by 4
	\else	\ifnum	\count 2 < -32767
		\then	\divide \count 2 by 4
			\divide \count 4 by 4
		\else
		\fi
	\fi
	\multiply \count 0 by \count 2
	\divide \count 0 by \count 4
	\xdef \product {#1 = \the \count 0 \internal@nits}%
	\aftergroup \product
       }}

\def\r@duce{\ifdim\dimen0 > 90\r@dian \then   % sin(x) = sin(180-x)
		\multiply\dimen0 by -1
		\advance\dimen0 by 180\r@dian
		\r@duce
	    \else \ifdim\dimen0 < -90\r@dian \then  % sin(x) = sin(360+x)
		\advance\dimen0 by 360\r@dian
		\r@duce
		\fi
	    \fi}

\def\Sine#1%
       {{%
	\dimen 0 = #1 \r@dian
	\r@duce
	\ifdim\dimen0 = -90\r@dian \then
	   \dimen4 = -1\r@dian
	   \c@mputefalse
	\fi
	\ifdim\dimen0 = 90\r@dian \then
	   \dimen4 = 1\r@dian
	   \c@mputefalse
	\fi
	\ifdim\dimen0 = 0\r@dian \then
	   \dimen4 = 0\r@dian
	   \c@mputefalse
	\fi
	\ifc@mpute \then
		\divide\dimen0 by 180
		\dimen0=3.141592654\dimen0
		\dimen 2 = 3.1415926535897963\r@dian %%% a well-known constant
		\divide\dimen 2 by 2 %%% we only deal with -pi/2 : pi/2
		\Mess@ge {Sin: calculating Sin of \nodimen 0}%
		\count 0 = 1 %%% see power-series expansion for sine
		\dimen 2 = 1 \r@dian %%% ditto
		\dimen 4 = 0 \r@dian %%% ditto
		\loop
			\ifnum	\dimen 2 = 0 %%% then we've done
			\then	\stillc@nvergingfalse 
			\else	\stillc@nvergingtrue
			\fi
			\ifstillc@nverging %%% then calculate next term
			\then	\term {\count 0} {\dimen 0} {\dimen 2}%
				\advance \count 0 by 2
				\count 2 = \count 0
				\divide \count 2 by 2
				\ifodd	\count 2 %%% signs alternate
				\then	\advance \dimen 4 by \dimen 2
				\else	\advance \dimen 4 by -\dimen 2
				\fi
		\repeat
	\fi		
			\xdef \sine {\nodimen 4}%
       }}

\def\Cosine#1{\ifx\sine\UnDefined\edef\Savesine{\relax}\else
		             \edef\Savesine{\sine}\fi
	{\dimen0=#1\r@dian\multiply\dimen0 by -1
	 \advance\dimen0 by 90\r@dian
	 \Sine{\nodimen 0}
	 \xdef\cosine{\sine}
	 \xdef\sine{\Savesine}}}	      
\def\psdraft{
	\def\@psdraft{0}
}
\def\psfull{
	\def\@psdraft{100}
}

\psfull

\newif\if@draftbox
\def\psnodraftbox{
	\@draftboxfalse
}
\@draftboxtrue

\newif\if@prologfile
\newif\if@postlogfile
\def\pssilent{
	\@noisyfalse
}
\def\psnoisy{
	\@noisytrue
}
\psnoisy
\newif\if@bbllx
\newif\if@bblly
\newif\if@bburx
\newif\if@bbury
\newif\if@height
\newif\if@width
\newif\if@rheight
\newif\if@rwidth
\newif\if@angle
\newif\if@clip
\newif\if@verbose
\newif\if@scale
\def\@p@@sclip#1{\@cliptrue}

\def\@p@@sfile#1{\def\@p@sfile{null}%
	        \openin1=#1
		\ifeof1\closein1%
		       \openin1=\figurepath#1
			\ifeof1\typeout{Error, File #1 not found}
			   \if@bbllx\if@bblly\if@bburx\if@bbury% added 6/91 Rob Russell
			      \def\@p@sfile{#1}%
			   \fi\fi\fi\fi
			\else\closein1
			    \edef\@p@sfile{\figurepath#1}%
                        \fi%
		 \else\closein1%
		       \def\@p@sfile{#1}%
		 \fi}
\def\@p@@sfigure#1{\def\@p@sfile{null}%
	        \openin1=#1
		\ifeof1\closein1%
		       \openin1=\figurepath#1
			\ifeof1\typeout{Error, File #1 not found}
			   \if@bbllx\if@bblly\if@bburx\if@bbury% added 6/91 Rob Russell
			      \def\@p@sfile{#1}%
			   \fi\fi\fi\fi
			\else\closein1
			    \def\@p@sfile{\figurepath#1}%
                        \fi%
		 \else\closein1%
		       \def\@p@sfile{#1}%
		 \fi}

\def\@p@@sbbllx#1{
		\@bbllxtrue
		\dimen100=#1
		\edef\@p@sbbllx{\number\dimen100}
}
\def\@p@@sbblly#1{
		\@bbllytrue
		\dimen100=#1
		\edef\@p@sbblly{\number\dimen100}
}
\def\@p@@sbburx#1{
		\@bburxtrue
		\dimen100=#1
		\edef\@p@sbburx{\number\dimen100}
}
\def\@p@@sbbury#1{
		\@bburytrue
		\dimen100=#1
		\edef\@p@sbbury{\number\dimen100}
}
\def\@p@@sheight#1{
		\@heighttrue
		\dimen100=#1
   		\edef\@p@sheight{\number\dimen100}
}
\def\@p@@swidth#1{
		\@widthtrue
		\dimen100=#1
		\edef\@p@swidth{\number\dimen100}
}
\def\@p@@srheight#1{
		\@rheighttrue
		\dimen100=#1
		\edef\@p@srheight{\number\dimen100}
}
\def\@p@@srwidth#1{
		\@rwidthtrue
		\dimen100=#1
		\edef\@p@srwidth{\number\dimen100}
}
\def\@p@@sangle#1{
		\@angletrue
		\edef\@p@sangle{#1} %\number\dimen100}
}
\def\@p@@ssilent#1{ 
		\@verbosefalse
}
\def\@p@@sscale#1{
		\def\@p@scale{#1}
		\@scaletrue
}
\def\@p@@sprolog#1{\@prologfiletrue\def\@prologfileval{#1}}
\def\@p@@spostlog#1{\@postlogfiletrue\def\@postlogfileval{#1}}
\def\@cs@name#1{\csname #1\endcsname}
\def\@setparms#1=#2,{\@cs@name{@p@@s#1}{#2}}
\def\ps@init@parms{
		\@bbllxfalse \@bbllyfalse
		\@bburxfalse \@bburyfalse
		\@heightfalse \@widthfalse
		\@rheightfalse \@rwidthfalse
		\@scalefalse
		\def\@p@sbbllx{}\def\@p@sbblly{}
		\def\@p@sbburx{}\def\@p@sbbury{}
		\def\@p@sheight{}\def\@p@swidth{}
		\def\@p@srheight{}\def\@p@srwidth{}
		\def\@p@sangle{0}
		\def\@p@sfile{}
		\def\@p@scost{10}
		\def\@sc{}
		\@prologfilefalse
		\@postlogfilefalse
		\@clipfalse
		\if@noisy
			\@verbosetrue
		\else
			\@verbosefalse
		\fi
}
\def\parse@ps@parms#1{
	 	\@psdo\@psfiga:=#1\do
		   {\expandafter\@setparms\@psfiga,}}
\newif\ifno@bb
\def\bb@missing{
	\if@verbose{
		\typeout{psfig: searching \@p@sfile \space  for bounding box}
	}\fi
	\no@bbtrue
	\epsf@getbb{\@p@sfile}
        \ifno@bb \else \bb@cull\epsf@llx\epsf@lly\epsf@urx\epsf@ury\fi
}	
\def\bb@cull#1#2#3#4{
	\dimen100=#1 bp\edef\@p@sbbllx{\number\dimen100}
	\dimen100=#2 bp\edef\@p@sbblly{\number\dimen100}
	\dimen100=#3 bp\edef\@p@sbburx{\number\dimen100}
	\dimen100=#4 bp\edef\@p@sbbury{\number\dimen100}
	\no@bbfalse
}

\newdimen\p@intvaluex
\newdimen\p@intvaluey
\newdimen\@ffsetvalue
\newdimen\x@ffsetvalue
\newdimen\y@ffsetvalue

\def\compute@offset#1#2{{\dimen0=#1 sp\dimen1=#2 sp
			\advance\dimen1 by -\dimen0
			\dimen1=\sine\dimen1
			\dimen0=\cosine\dimen1
			\ifdim\dimen0<0sp \dimen1=0sp \fi
			\global\@ffsetvalue=\dimen1}}

\def\rotate@#1#2{{\dimen0=#1 sp\dimen1=#2 sp
		  \global\p@intvaluex=\cosine\dimen0
		  \dimen3=\sine\dimen1
		  \global\advance\p@intvaluex by -\dimen3
		  \global\p@intvaluey=\sine\dimen0
		  \dimen3=\cosine\dimen1
		  \global\advance\p@intvaluey by \dimen3
		  }}
\def\compute@bb{
		\no@bbfalse
		\if@bbllx \else \no@bbtrue \fi
		\if@bblly \else \no@bbtrue \fi
		\if@bburx \else \no@bbtrue \fi
		\if@bbury \else \no@bbtrue \fi
		\ifno@bb \bb@missing \fi
		\ifno@bb \typeout{FATAL ERROR: no bb supplied or found}
			\no-bb-error
		\fi
		\if@angle 
			\Sine{\@p@sangle}\Cosine{\@p@sangle}
			\compute@offset{\@p@sbblly}{\@p@sbbury}
			\x@ffsetvalue=\@ffsetvalue
			\compute@offset{\@p@sbburx}{\@p@sbbllx}
			\y@ffsetvalue=\@ffsetvalue

			\rotate@{\@p@sbbllx}{\@p@sbblly}
			\advance\p@intvaluex by -\x@ffsetvalue
			\advance\p@intvaluey by -\y@ffsetvalue
			\edef\@p@sbbllx{\number\p@intvaluex}
			\edef\@p@sbblly{\number\p@intvaluey}

			\rotate@{\@p@sbburx}{\@p@sbbury}
			\advance\p@intvaluex by \x@ffsetvalue
			\advance\p@intvaluey by \y@ffsetvalue
			\edef\@p@sbburx{\number\p@intvaluex}
			\edef\@p@sbbury{\number\p@intvaluey}
			{
			 \count0=\@p@sbbllx \count1=\@p@sbblly
		 	 \count2=\@p@sbburx \count3=\@p@sbbury
			 \dimen0=\@p@sbbllx sp\dimen1=\@p@sbblly sp
		 	 \dimen2=\@p@sbburx sp\dimen3=\@p@sbbury sp
			 \dimen203=\dimen2 \advance\dimen203 by -\dimen0
			 \dimen204=\dimen3 \advance\dimen204 by -\dimen1
			 \ifdim\dimen203<0sp 
			      \count203=\count2 \count2=\count0 
			      \count0=\count203 
			      \global\edef\@p@sbbllx{\number\count0}
			      \global\edef\@p@sbburx{\number\count2}
			 \fi
			 \ifdim\dimen204<0sp 
			       \count204=\count3
			       \count3=\count1
			       \count1=\count204
			       \global\edef\@p@sbblly{\number\count1}
			       \global\edef\@p@sbbury{\number\count3}
			 \fi
			}
		\fi
		\count203=\@p@sbburx
		\count204=\@p@sbbury
		\advance\count203 by -\@p@sbbllx
		\advance\count204 by -\@p@sbblly
		\edef\@bbw{\number\count203}
		\edef\@bbh{\number\count204}
}
\def\in@hundreds#1#2#3{\count240=#2 \count241=#3
		     \count100=\count240	% 100 is first digit #2/#3
		     \divide\count100 by \count241
		     \count101=\count100
		     \multiply\count101 by \count241
		     \advance\count240 by -\count101
		     \multiply\count240 by 10
		     \count101=\count240	%101 is second digit of #2/#3
		     \divide\count101 by \count241
		     \count102=\count101
		     \multiply\count102 by \count241
		     \advance\count240 by -\count102
		     \multiply\count240 by 10
		     \count102=\count240	% 102 is the third digit
		     \divide\count102 by \count241
		     \count200=#1\count205=0
		     \count201=\count200
			\multiply\count201 by \count100
		 	\advance\count205 by \count201
		     \count201=\count200
			\divide\count201 by 10
			\multiply\count201 by \count101
			\advance\count205 by \count201
		     \count201=\count200
			\divide\count201 by 100
			\multiply\count201 by \count102
			\advance\count205 by \count201
		     \edef\@result{\number\count205}
}
\def\@ScaleInHundreds#1{
		\in@hundreds{#1}{\@p@scale}{100}
		\edef#1{\@result}
}
\def\compute@wfromh{
		\in@hundreds{\@p@sheight}{\@bbw}{\@bbh}
		\edef\@p@swidth{\@result}
}
\def\compute@hfromw{
		\in@hundreds{\@p@swidth}{\@bbh}{\@bbw}
		\edef\@p@sheight{\@result}
}
\def\compute@handw{
		\if@height 
			\if@width
			\else
				\compute@wfromh
			\fi
		\else 
			\if@width
				\compute@hfromw
			\else
				\edef\@p@sheight{\@bbh}
				\edef\@p@swidth{\@bbw}
			\fi
		\fi
}
\def\compute@resv{
		\if@rheight \else \edef\@p@srheight{\@p@sheight} \fi
		\if@rwidth \else \edef\@p@srwidth{\@p@swidth} \fi
}
\def\compute@sizes{
	\compute@bb
	\compute@handw
	\compute@resv
}
\def\psfig#1{\vbox {
	\ps@init@parms
	\parse@ps@parms{#1}
	\compute@sizes
	\if@scale
                \if@verbose
                        \typeout{psfig: scaling by \@p@scale}
                \fi
                \@ScaleInHundreds{\@p@swidth}
                \@ScaleInHundreds{\@p@sheight}
                \@ScaleInHundreds{\@p@srwidth}
                \@ScaleInHundreds{\@p@srheight}
        \fi
	\ifnum\@p@scost<\@psdraft{
		\if@verbose{
			\typeout{psfig: including \@p@sfile \space }
		}\fi
		\special{ps::[begin] 	\@p@swidth \space \@p@sheight \space
				\@p@sbbllx \space \@p@sbblly \space
				\@p@sbburx \space \@p@sbbury \space
				startTexFig \space }
		\if@angle
			\special {ps:: \@p@sangle \space rotate \space} 
		\fi
		\if@clip{
			\if@verbose{
				\typeout{(clip)}
			}\fi
			\special{ps:: doclip \space }
		}\fi
		\if@prologfile
		    \special{ps: plotfile \@prologfileval \space } \fi
		\special{ps: plotfile \@p@sfile \space }
		\if@postlogfile
		    \special{ps: plotfile \@postlogfileval \space } \fi
		\special{ps::[end] endTexFig \space }
		\vbox to \@p@srheight true sp{
			\hbox to \@p@srwidth true sp{
				\hss
			}
		\vss
		}
	}\else{
		\if@draftbox{		
			\hbox{\fbox{\vbox to \@p@srheight true sp{
			\vss
			\hbox to \@p@srwidth true sp{ \hss \@p@sfile \hss }
			\vss
			}}}
		}\else{
			\vbox to \@p@srheight true sp{
			\vss
			\hbox to \@p@srwidth true sp{\hss}
			\vss
			}
		}\fi

	}\fi
}}
\def\psglobal{\typeout{psfig: PSGLOBAL is OBSOLETE; use psprint -m instead}}
\psfigRestoreAt